\documentclass[traditabstract]{aa}  
\usepackage{natbib}
\usepackage{graphicx}
\begin{document}

   \title{LBT observations of the HR 8799 planetary system
          \thanks{The LBT is an international collaboration among institutions in the United States, Italy and Germany. LBT Corporation partners are: The University of Arizona on behalf of the Arizona university system; Istituto Nazionale di Astrofisica, Italy; LBT Beteiligungsgesellschaft, Germany, representing the Max-Planck Society, the Astrophysical Institute Potsdam, and Heidelberg University; The Ohio State University, and The Research Corporation, on behalf of The University of Notre Dame, University of Minnesota and University of Virginia.}}

\subtitle{First detection of HR8799e in H band}

\titlerunning{LBT observations of the HR 8799 planetary system}

   \author{S. Esposito
          \inst{1},
           D. Mesa
           \inst{2},
           A. Skemer
           \inst{3},
           C. Arcidiacono
           \inst{1,4},  
           R.U. Claudi
          \inst{2},
           S. Desidera
          \inst{2},
           R. Gratton
           \inst{2},
           F. Mannucci
           \inst{1},
          F. Marzari
          \inst{5},
           E. Masciadri
           \inst{1},
           L. Close
           \inst{3},
           P. Hinz
           \inst{3},
           C. Kulesa
           \inst{3},
           D. McCarthy  
           \inst{3},
           J. Males  
           \inst{3},         
        G. Agapito
         \inst{1},
         J. Argomedo
         \inst{1},
         K. Boutsia
         \inst{6,7}
         R. Briguglio
          \inst{1},
          G. Brusa
         \inst{6},
         L. Busoni
         \inst{1},
         G. Cresci
          \inst{1},
           L. Fini
          \inst{1},
           A. Fontana
          \inst{7},
         J.C. Guerra
         \inst{6},
         J.M. Hill
          \inst{6},
        D. Miller
         \inst{6},
         D. Paris
         \inst{7},
          E. Pinna
          \inst{1},
          A. Puglisi
          \inst{1},
         F. Quiros-Pacheco
         \inst{1},
           A. Riccardi
           \inst{1},
         P. Stefanini
         \inst{1},
         V. Testa
           \inst{7},
          M. Xompero
         \inst{1},    
         C. Woodward       
          \inst{8}}

   \authorrunning{S. Esposito et al.}

   \offprints{S. Esposito,  \\
              \email{esposito@arcetri.astro.it} }

   \institute{INAF -- Osservatorio Astrofisico di Arcetri
              L.go E. Fermi 5
              50125 Firenze, Italy
                \and 
              INAF -- Osservatorio Astronomico di Padova,  
              Vicolo dell' Osservatorio 5, I-35122, Padova, Italy
             \and
              Steward Observatory, Department of Astronomy, 
              University of Arizona, Tucson, AZ 85721, USA
             \and
              INAF -- Osservatorio Astronomico di Bologna, Via Ranzani 1, 40127, Bologna, Italy
            \and 
             Dipartimento di Fisica e Astronomia -- Universit\'a di Padova, Via Marzolo 8,
             Padova, Italy
             \and
              LBT Observatory, Univ. of Arizona, 933 North Cherry Ave., Tucson AZ 85721, USA 
             \and
             INAF -- Osservatorio Astronomico di Roma, Via Frascati 33, 00040 Monteporzio (RM), Italy
             \and
             Minnesota Institute of Astrophysics, University of Minnesota, Minneapolis, MN 5455, USA
              }

 \date{Received  / Accepted }

\abstract{We have performed $H$ and $K_{S}$ band observations of the planetary system around HR 8799 
using the new AO system at the Large Binocular Telescope and the PISCES Camera.
The excellent instrument performance (Strehl ratios up to 80\% in $H$ band) enabled
detection of the inner planet HR8799$e$ in the H band for the first time.
The $H$ and $K_{S}$  magnitudes of HR8799e are similar to those of planets $c$ and $d$, with planet $e$
slightly brighter. Therefore,  HR8799$e$ is likely slightly more massive than $c$ and $d$.
We also explored possible orbital configurations and their orbital stability.
We confirm that the orbits of planets $b$, $c$ and $e$ are consistent with being circular and coplanar;
planet $d$ should have either an orbital eccentricity of about 0.1 or be non-coplanar with respect to
$b$ and $c$. Planet $e$ can not be in circular and coplanar orbit in a 4:2:1 mean motion resonances with $c$ and $d$, while
coplanar and circular orbits are allowed for a 5:2 resonance.
The analysis of dynamical stability shows that the system is highly unstable or chaotic when planetary 
masses of about 5~$M_{J}$ for $b$ and 7~$M_{J}$ for the other planets are adopted. 
Significant regions of dynamical stability for timescales of tens of Myr are found when adopting
planetary masses of about 3.5, 5, 5, and 5 $M_{J}$ for HR 8799 $b$, $c$, $d$, and $e$ respectively.
These masses are below the current estimates based on the stellar age (30 Myr) and theoretical models
of substellar objects.}


   \maketitle
%

\section{Introduction}
\label{s:intro}

The planetary system around \object{HR8799} represents a unique laboratory to constrain the
physical properties of massive giant planets, to study the architecture 
of a crowded planetary system, and the link between planets and debris belts.

Three planets (\object{HR 8799 $b$}, $c$ and $d$) were discovered by \cite{2008Sci...322.1348M}, 
at a projected separation of about 24, 38, and 68 AU, followed by the detection of an 
inner planet (\object{HR8799 $e$}) at about 15 AU \citep{2010Natur.468.1080M}.
The system is completed by three debris disk components, a belt of warm dust
($T \sim 150$~K) between about 6 to 10 AU, a broad belt of cold dust ($T \sim 45$~K)
between 90 to 300 AU,
whose inner edge is probably defined by the interactions with the outer planet, and
an extended halo of small grains up to 1000 AU \citep{2009ApJ...705..314S}.
The belt of cold dust at about 100 AU have been spatially resolved at $70 \mu m$ using Spitzer
\citep{2009ApJ...705..314S}.
The central star is an A5 star located at 39.4 pc from the Sun \citep{2007A&A...474..653V}, characterized by 
$\lambda$ Boo-like abundances anomalies and $\gamma$ Doradus pulsations \citep{1999AJ....118.2993G}.

The architecture of the HR8799 system, with its four giant planets and two belts, resembles that of our 
Solar system, especially when the two systems
are plotted against the equilibrium temperature at various distances from
the central star, taking the higher luminosity of HR8799 compared
to the Sun into account \citep{2010Natur.468.1080M}. However, the planets around HR 8799 are
much more massive than those in the Solar System.

The discovery of this planetary system prompted several investigations focused
mostly on the physical properties of the planets \citep[e.g.][]{2010ApJ...723..850B,2011ApJ...729..128C}
and the dynamical stability of the planetary system \citep[e.g.][]{2009MNRAS.397L..16G,2010ApJ...710.1408F}.

The masses of the planets have not been determined dynamically and estimates are 
therefore derived from the stellar age of HR8799 and theoretical models.
\cite{2008Sci...322.1348M} estimated age limits between 30 to 160 Myr, from
the position of HR8799 on HR diagram. 
\cite{2010Natur.468.1080M} narrowed the plausible age range to 30-60 Myr
(with preference for the younger value) by classifying HR 8799 as a probable member
of the \object{Columba moving group} \citep{2008hsf2.book..757T}.
The association of HR 8799 with Columba was questioned by \cite{2010ApJ...716..417H}, 
who noted that the closest approach to the centroid of Columba moving group
was $\sim 58$~pc 27 Myr ago (this was considered too large for a direct link) and that
the size of this and other young groups might be too large to have a common origin.
Furthermore, available models of substellar objects suffer of significant uncertainties
especially at young ages \citep{2002A&A...382..563B}, leaving some ambiguity
on the planet masses as derived from magnitudes or colors even at fixed age.

On the other hand, studies of dynamical stability of the system showed
that, for masses above $20~M_{J}$, it is basically impossible to find
orbital configurations compatible with the astrometric data and that are, at the same time,
stable for the age of the system \citep{2010ApJ...721L.199M}.
The packed configuration of the system then favour the lowest planetary masses
(corresponding to the  youngest ages for the system), i.e. about 
$5~M_{J}$ for the outer planet and $7~M_{J}$ for the other three.
Most of these studies were based on a 3-planet system architecture
and should be extended including the fourth planet.
The continuation of the astrometric monitoring and its ``extension'' to 
the past by identification of some of the planets in improved reanalysis of 
past data \citep{2011ApJ...741...55S} is expected to provide tighter constrain on
both the planetary orbits and masses.

The development of new instrumentation for imaging of giant planets at small
separation is crucial for a further understanding of the system.
Enhanced Strehl ratios allow to extend the detection space
of the inner planets to additional wavelengths and then to better
characterize their physical properties. In addition they allow to improve
the accuracy of astrometric and photometric measurements thanks to the reduced speckle noise
and the enhanced contrast of the planet PSF.

The new adaptive optics system (FLAO) of the Large Binocular Telescope (LBT)
has achieved, since its commissioning, 
unprecedented performance with Strehl Ratios higher than 80\% in H band
\citep{2010ApOpt..49G.174E}.
Images of HR8799 system with such an instrumentation allowed us to detect
for the first time the inner planet (HR8799e) in H band, enabling a more robust characterization of
this planetary system.
The present paper describes  the instrumentation and the observed procedures
adopted to achieve this result, the data analysis procedures. Finally, it
discusses the results both in terms of physical characterization of the planets
and of the dynamical architecture of the system.
A companion paper, \cite{skemer12}, presents 3.3$\mu m$ photometry 
of the HR 8799 system obtained using the Large Binocular
Telescope Interferometer (LBTI) and the FLAO system, together with
the independent analysis of the H band data discussed here and a comparison of
the spectral energy distributions of the four planets with 
a variety of models including a new set of mixed-cloud models.

\section{Observations}
\label{s:obs}

\subsection{The LBT Adaptive Optics system}
\label{s:flao}

The Large Binocular Telescope (LBT) is a unique telescope featuring two co-mounted optical 
trains with 8.4m primary mirrors~\citep{2010ApOpt..49D.115H}. The First-Light Adaptive Optics 
system (FLAO) of the LBT takes advantage of two innovative key components, namely an adaptive 
secondary mirror with 672 actuators and a high-order pyramid wave-front sensor with a maximum 
pupil sampling of 30$\times$30 subapertures~\citep{2010ApOpt..49G.174E}.  
FLAO\#1 system is located on the right telescope bent Gregorian focal station
and controls the right secondary mirror. The WFS is mounted on rotating bearing
that allows to compensate sky rotation: it receives the visible light reflected by
a dichroic mirror, which transmits the infrared to the scientific instrument (in our case Pisces).
The commissioning of the instrument was completed in the winter of 2011, 
including a period of Science Demonstration Time (SDT) supplying a corrected wavefront for the 
PISCES imaging Near Infrared (NIR) Camera \citep{2001PASP..113..353M}. The full adaptive optics 
imaging and spectroscopic channel will be completed when the LUCIFER \citep{2000SPIE.4008..767M} 
instrument will be installed on the NIR focal plane corrected by the FLAO\#1.

Since the initial phases of the on-sky commissioning, the FLAO\#1 system reached performances 
never achieved before on large ground-based optical telescopes. Images with 40mas resolution 
and Strehl Ratios higher than 80\% were obtained in H band (1.6 $\mu$m). 
The images show a ratio between the intensity at 0.4 arcsec and the
central peak larger than $10^{-4}$ \citep{2011SPIE.8149E...1E}.

\subsection{PISCES Camera}
\label{s:pisces}

The observations presented here used the 1-2.5 $\mu m$ camera PISCES
\citep{2001PASP..113..353M} at a bent-Gregorian focus of one 8.4 m primary
mirror of the Large Binocular Telescope (LBT). Internal optics, cooled
to 77 K, reimaged the f/15 focal plane onto a HAWAII-1 detector (1024
pixel square) at f/23.5, yielding a scale of $19.31\pm0.03$ mas/px
with a field-of-view of 19.7 arcsec on a side \citep{close12}. This
scale critically samples the diffraction-limit in the H-band ($\lambda$/D
= 40.5 mas). A cold pupil stop, nearly conjugated to the adaptive
secondary mirror, shielded unwanted background radiation.  A dichroic
beamsplitter located in the converging beam ahead of the camera
directed visible light ($<0.95~\mu m$) onto the wavefront sensor unit. Images from PISCES are
obtained from double-correlated sampling with a read-noise of ~20
electrons. 
Further details on the performances of the PISCES camera coupled with the LBT
AO system are described in \cite{guerra12}.

\subsection{Observing strategy}
\label{s:obstrat}

Our observations of HR~8799 were obtained with the
Large Binocular Telescope in two different filters: H~band on 2011 October 16 UT, 
within the LBT PISCES+AO Science Verification Time (SV), and 
in Ks band on 2011 November 09 UT during the Science Demonstration Time 
(SDT). The summary of the observing setup and of the observing conditions for
these two epochs is reported in Table~\ref{tabsetup}.

\begin{table}
\caption{Summary of the observing setup and the observing conditions for HR8799. \label{tabsetup}}
\begin{center}
\begin{tabular}{c c c}
\hline\hline
      &      H band filter    &  Ks band filter      \\ \hline
  Observation date (UT)   &    2011 Oct. 16    &    2011 Nov. 09   \\  
  Number of images   &    901    &     328    \\
  Exposure time      &    2 s    &    2 s     \\
  Total integration time  &    30 min    &   11 min               \\
  Total field rotation  &  $89.6^{\circ}$  &     $36.0^{\circ}$    \\   
  Seeing      &    0.93 arcsec     &    1.0 arcsec     \\  \hline

\end{tabular}
\end{center}
\end{table}

The image rotator was stopped
to enable angular differential
imaging \citep{2006ApJ...641..556M}. During the observations in H band, 
in order to minimize the dynamic effect of quasi-static speckles, we concentrated the observations close to the meridian 
passage of the star achieving the largest angular coverage in the small time frame available.
However, this strategy was not used during the observation in Ks band because observations started slightly after the meridian passage due to technical problems.
The seeing was measured with the LBT-DIMM pointing to the same direction of the scientific target on sky.

We observed HR8799 saturating the inner region in both runs (i.e. in both filters H and
Ks) at  radii closer to 160 mas.  At radii larger than 160 mas, the 
planets were observed with the stellar halo within the linear regime of the camera (that is below 8000-10000 counts).
We collected calibration images to compute differential sky flat-field at the sunset and at 
the sunrise of the observing night.

\section{Data analysis procedure}
\label{s:analysis}

Initial image processing corrects all raw images for electronic cross-talk between the quadrants 
in the detector using Corquad, an IRAF\footnote{IRAF is distributed by the National Optical 
Astronomy Observatory, which is operated by the Association of Universities for Research in 
Astronomy (AURA) under cooperative agreement with the National Science Foundation.} task developed 
for this purpose\footnote{Corquad is available at \tt http://aries.as.arizona.edu/}.
The cross-talk coefficients were updated after final electronic set-up at LBT as described in \cite{guerra12}
After this, data analysis aiming to  achieve  the  best  possible  contrast  was  performed  using  two
independent pipelines and  data-reduction strategies called, hereafter
"A" and  "B" methods developed respectively at  the Padova Observatory
and at Steward Observatory. Both pipelines achieved coherent results.

\subsection{Pipeline A}\label{s:pipelinea}

The data analysis is composed of different steps that were implemented using IDL routines prepared for this 
purpose. The adopted procedure represents an optimization of data analysis routines prepared for similar datasets 
obtained with NACO at VLT \citep{2010lyot.confE..23C}. The first step, which is critical because of heavy 
saturation, concerns the identification of the center of the star in each image.
Accurate frame-to-frame relative positions were obtained by performing a cross-correlation (CC)
between one image chosen as reference (the first image in the dataset) and all the other ones  
The peak of the CC represents the
shift between the center position of the two images. The position of the maximum was 
obtained by means of a 2D Gaussian fit on this peak.This was found to be more
precise than the Gaussian fit of the image.
Much more difficult is to derive the position of the center of the star, 
with respect to that of faint planet images, that are not
detectable in individual images, We proceeded as follows.
First, the absolute position was obtained by finding the center of the stellar
image after heavily smoothing using a 2D Gaussian fit. This procedure was repeated
for every image. The results were then corrected for the relative frame-to-frame
positions determined above, and then averaged. We found that on average the centers
have an offset of about 0.5 pixels in both coordinates with respect to the value
determined from the first image. We apply this correction to our star center
position. The r.m.s. scatter for individual images is 0.31 pixels in both coordinates.
Assuming that errors in the individual determinations
are independent, a very small error of ~0.01 pixels
($\sim 0.2$~mas) is derived for the H-band data, and about twice this value
for those in the K-band. However, systematic errors due to asymmetries of the
PSF's are likely much larger. A rough estimate can be obtained
by determining the star center in a slighly different way.
Rather than heavily smoothing the profiles,
we replaced pixel values in the saturated region with a constant value close to the maximum
of the unsaturated pixels. The resulting center positions differ
systematically by $0.116\pm 0.005$~pixels with respect to the adopted ones.
Since both procedure looks fairly legitimate, we conclude that our star position
may have systematic errors as large as 0.1 pixels and likely more, as both procedures
take into account only the outer regions of the PSF, and may underestimate
asymmetries of these with respect to the core of the PSF, which is the only part
that can be detected for the planets. 
Therefore, conservatively we adopt a larger error of 0.5 pixels in our discussion.

In the second step, a 2D stellar profile is subtracted from each single image. The aim is 
to reduce the strong signal gradients present in the image, improving effectiveness of the following 
filtering procedures. This was done by subtracting to the values measured in each pixels
the median of the counts in annuli one pixel wide at different separations from the stellar center. After this, 
a low-pass filtering was applied to each image to eliminate bad and hot pixels. The next 
step was an high-pass filtering of the images. This was done by subtracting from each pixel the median of 
a sub-image composed by $n\times n$\ pixels around it. The value of $n$\ changes according to the distance 
from the center of the star and it can be optimized in such a way to obtain the best final result. However, 
to reduce the possibility to self-subtract a possible companion, the median was performed without 
considering the central pixels of the sub-image.

Finally, we implemented our version of the angular differential imaging (ADI) method ~\citep{2006ApJ...641..556M}. We first
located the center of the star of each image in the same position. Then, for each image, we selected a sub-sample of  
images in such a way that  their rotation with respect to the reference image was not too small, 
to avoid to self-subtract possible companion objects, and not too large, otherwise the speckle pattern would change 
too much and the image subtraction is no more effective in the speckle noise subtraction. The best criteria for
image selection are different at different separations from the central star, so  we repeated this part 
of the procedure using different sets of images optimized for different separations. We then evaluated a median image for
each sub-sample of images giving larger weights to images taken at shorter time lapses and the resulting image 
was then subtracted from the reference one. This subtracted image was then rotated by the 
appropriate value, given by the parallactic angle as reported in the image header. To this angle, however, 
we had to add a further rotation to obtain a proper absolute orientation of each image. The calibration of 
this position angle is preliminary.
However, an a posteriori test of the accuracy of the zero point correction to true north is given by the 
consistency of our astrometry with the predictions of the orbit of planet b by Soummer et al. (2011). 
This test indicates that the zero point is not in error by more than 0.3 degrees, in agreement with 
the error in the position of the true north by Close et al. (2012). This procedure is repeated for 
each image of the datacube. 
After this, the companion objects were in the same position in every image. We searched for such
companions on the median of this datacube.

For H band series, the best results in terms of the image S/N ratio for photometry at the observed angular 
separation of HD8799$e$ from the primary star are achieved using 
only the frames taken closer to meridian passage (216 frames corresponding 
to $\pm 15^{\circ}$, for a total integration time of 7 minutes). This is due to a better subtraction
of speckle for larger angular velocity and shorter time baseline with respect to the full dataset.
At larger separation, where speckle noise is less critical, the use
of a larger dataset provides a slightly better S/N due to the larger photon flux.
The results presented in Sect.~\ref{s:results} are based on the subset of the 216 images taken
close to meridian passage.

For the case of the $K_{S}$ band we used all the images (apart three
of very poor quality) because it was not possible to observe the star during meridian 
transit due to technical problems.

\subsection{Pipeline B}

The H-band data were independently reduced with a LOCI-based pipeline
\citep[Locally Optimized Combination of Images;][]{2007ApJ...660..770L},
which is hereafter referred to as Pipeline B.  
Pipeline B begins by flat/dark/distortion correcting the
first 500 images (rather than 216 images for pipeline A), before the
natural seeing worsened.  Processed images were aligned by
cross-correlation. The stellar profile was subtracted from each image,
using an azimuthal average, and quasi-static speckles were suppressed
by subtracting the median of the full set of images from each image.
Through this step, pipeline A and pipeline B are mostly similar, other
than the fact that pipeline B uses more data frames, and pipeline A
includes an additional high-pass/low-pass filter (they are also
independent implementations).  After these (fairly standard) ADI
steps, pipeline B uses LOCI to further suppress the noise of
quasi-static speckle residuals.
Pipeline B's results are statistically consistent with the results of Pipeline A.  
An in-depth description of Pipeline B, and its photometric results are
presented in a companion paper, \cite{skemer12}.  In the context
of this work, it is important to highlight that independent pipelines have produced
similar results given that they are being used with a new instrumental
setup. Some discrepancies are present in astrometric results between the two pipelines. This
should probably arise from the different images centering procedure that, as said in the 
previous Paragraph, is probably the more tricky step of our data reduction procedures.
 
\subsection{Astrometric calibration and distortion correction}\label{s:astrocal}

A laser cut sieve mask was used in laboratory to derive the distortion correction 
for the PISCES camera \citep{guerra12}. The coefficients of the polynomial computed in this way, reported in Appendix~\ref{app:distortion}, have a 0.6-pixel accuracy in one $\sigma$ error. This error does not affect 
the data reduction since it produces negligible correction (less than 0.1 pixels)
on distances of the order of the separation of HR8799 planets.

The absolute plate scale (19.274 mas/pix) was taken from \cite{close12} observations of the Orion Trapezium field.
Preliminary analysis of on sky observations of galactic Globular Cluster compared to HST calibrated ones 
further give support to the adopted value. 
Following \cite{close12} we also adopted a true north correction of $0.9\pm0.3^{\circ}$.

\section{Results}
\label{s:results}

\begin{figure}
\begin{center}
\includegraphics[width=8.0cm]{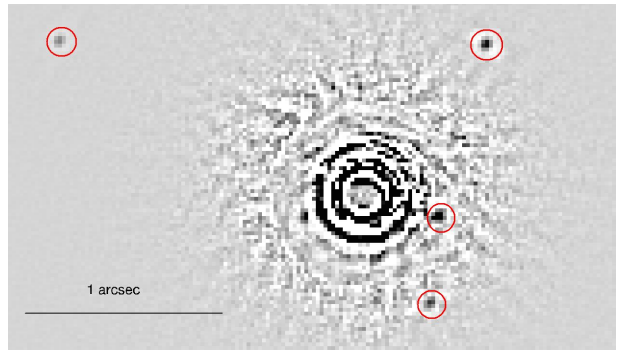}
\caption{Final image obtained from the H band data. \label{hbandfinal}}
\end{center}
\end{figure}

\begin{figure}
\begin{center}
\includegraphics[width=8.0cm]{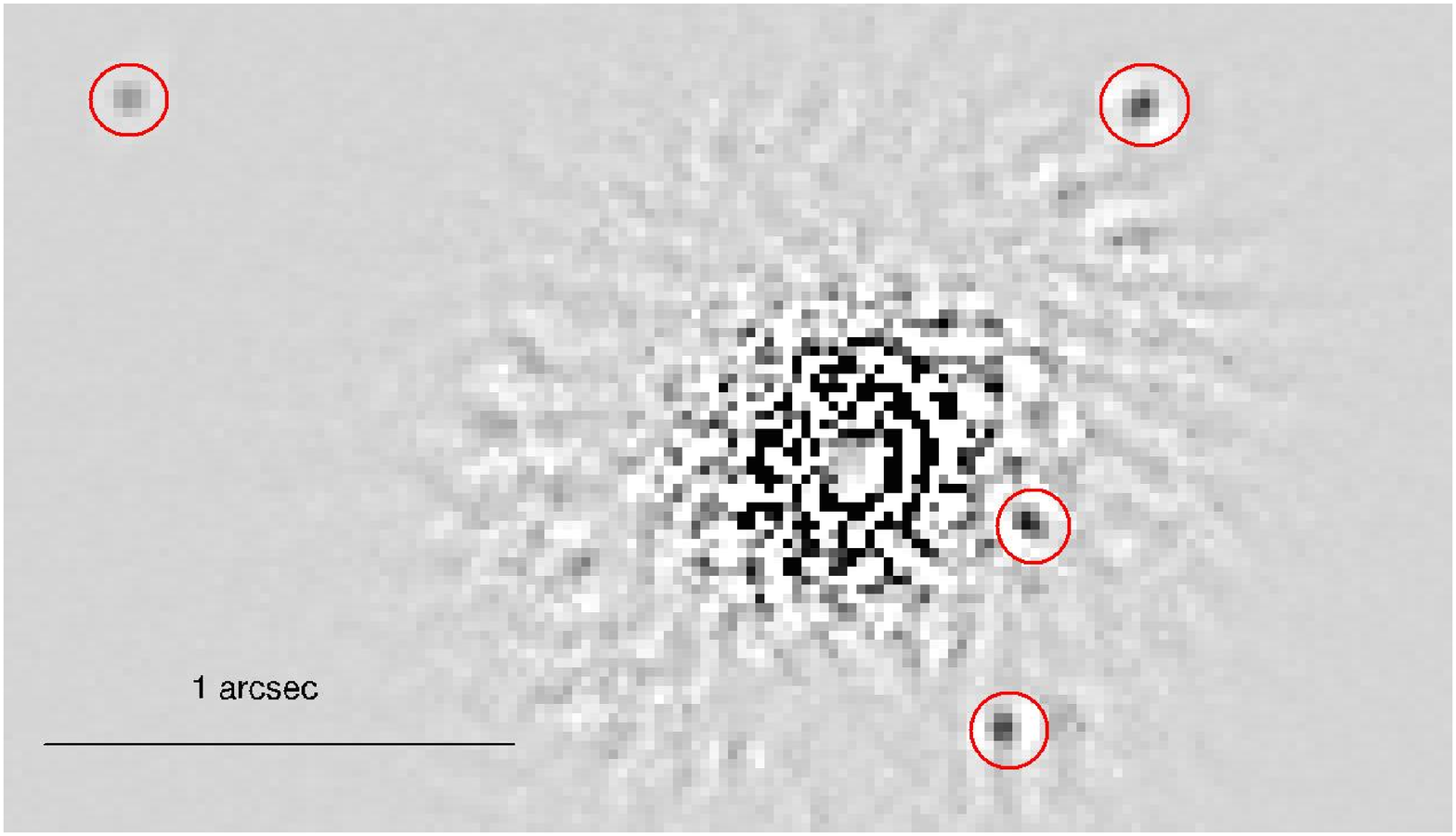}
\caption{Final image obtained from the K$_{S}$ band data. \label{kbandfinal}}
\end{center}
\end{figure}

In Figure~\ref{hbandfinal} and Figure~\ref{kbandfinal} we display the final best ADI image obtained using the H- and
K-band data, respectively\footnote{To reduce the noise visible in the final  image, a smoothed
image obtained using the IDL 'smooth' procedure was subtracted. Astrometry and photometry were performed on the original images.}. 
In both figures the four planets are clearly visible; for best identification
we marked their position with red circles. The FWHM of the planetary images are about 2.9 pixels,
slightly larger than expected value of diffraction peaks. 


We measured both position and luminosity of the four detected planets. 

\subsection{Astrometry}\label{sub:astrometry}
In Table~\ref{tabastrometryh} and in Table~\ref{tabastrometryk} we report our astrometric measurements for the four planets
around HR8799 in the H and in the Ks-band respectively. Table~\ref{taberrastro} gives the 
astrometric error budget. We considered several sources of errors: centring of the star, image
orientation (true north), scale distortion and planet centering errors (photometric errors).
These last were evaluated comparing the measured position of a number of simulated planets, 
inserted into the original images, with their original values. Since these fake planets were
inserted on the original images, these errors include artefacts due to data reduction and ADI, 
as well as the impact of speckles.

\begin{table}
\caption{Astrometry (measured with respect to the centroid of the star HR8799) obtained from the H-band 
data (epoch 2011.79). \label{tabastrometryh}}
\begin{center}
\begin{tabular}{c c c}
\hline\hline
  Planet   &      $\Delta$RA (arcsec)    &   $\Delta$Dec (arcsec)      \\ \hline
  b   &    1.579$\pm$0.011    &     0.734$\pm$0.011                \\
  c   &    -0.561$\pm$0.010   &    0.752 $\pm$0.010               \\
  d  &    -0.299$\pm$0.010    &   -0.563 $\pm$0.010               \\
  e  &    -0.326$\pm$0.011    &   -0.119 $\pm$0.011               \\   \hline

\end{tabular}
\end{center}
\end{table}

\begin{table}
\caption{Astrometry (measured with respect to the centroid of the star HR8799) obtained from the Ks-band data (epoch 2011.86). \label{tabastrometryk}}
\begin{center}
\begin{tabular}{c c c }
\hline\hline
  Planet   &      $\Delta$RA (arcsec)    &   $\Delta$Dec (arcsec)        \\ \hline
  b   &    1.546 $\pm$0.011   &     0.725$\pm$0.011                \\
  c   &    -0.578 $\pm$0.010  &    0.767$\pm$0.010                \\
  d  &    -0.320$\pm$0.010    &   -0.549$\pm$0.010                \\
  e  &    -0.382$\pm$0.011    &   -0.127$\pm$0.011                \\   \hline
\end{tabular}
\end{center}
\end{table}

\begin{table}
\caption{Astrometric error budgets (in mas) for the four HR8799 planets. \label{taberrastro}}
\begin{center}
\begin{tabular}{c c c c c c c}
\hline\hline
Error source      &  Error       &   b   &   c   &   d   &    e   &   Notes      \\ \hline
Star pos.         &  0.5 pix     &   9   &   9   &   9   &    9   & Same for all planets \\
True north        &  0.3 deg     &   6   &   4   &   2   &    1   & Error in pos. angle  \\
Scale             &  0.2\%       &   2   &   1   &   1   &    1   & Error in radial sep. \\
Distortion        &  0.15\%      &   2   &   2   &   2   &    2   & Both in x and in y  \\  
Phot. err.        &              &   0   &   0   &   1   &    5   & Both in x and in y \\ \hline
\end{tabular}
\end{center}
\end{table}

Final errors in our astrometry were obtained by combining quadratically uncertainties concerning star position, true north, scale, distortion and statistical error in photometry. 
The dominant term is star centering. Uncertainties in the image orientation give a significant
contribution for the most external planet while the photometric error is not negligible for the
planet e.

\subsection{Photometry} \label{sub:photometry}

Because of saturation of the center of the star and the lack of unsaturated exposures, 
we can not use the star magnitude as a reference to
evaluate the planet ones. 
For this reason we assumed, as a reference of our calculation, the magnitude of HR8799$b$ from \citet{2008Sci...322.1348M}
and we calculated the magnitude of the other planets by the ratio of planets counts. Pixels with separation $\leq$ from 
the position of each planet were used for 
the relative photometry.
We then repeated the same procedure assuming the same thing for HR8799$c$ and we made a mean between the two results to obtain 
the final result. Given that HR8799$b$ and HR8799$c$ are located at quite large distances from the center of 
the star (1.72 and 0.97 arcsec respectively), this assumption is reasonable because the images of the planets are 
weakly affected by speckles.
Table~\ref{tabphotometry} reports the absolute magnitude of the four exo-planets 
as well as the associated errors. 
The magnitudes are corrected for the self-subtraction effect introduced by the ADI procedure.
A set of errors (photometric errors, error due to the self-subtraction of the SF of the star and errors due 
to the filtering data-reduction procedure) were 
evaluated by inserting into the original images $\sim 10$\ template planets with the same counts and at the same 
separation of each of the four planets.  
The same data analysis procedure was then performed on these latter images. 
The standard deviation of the counts for planets at the same separation was then taken as the uncertainty 
on the photometry of each planet.
Our H band photometry is on average about 0.20 mag fainter than that by \cite{skemer12}, with a dispersion of 0.13 mag. 
This offset is primarily due to different choices for reference magnitude adopted for the outer 
planets, \citet{2008Sci...322.1348M} in our case and \citet{2009ApJ...705L.204M} for \citet{skemer12}, which
implies a systematic difference of 0.21 mag. 

Beside the agreement with the photometry by \citet{2008Sci...322.1348M,2010Natur.468.1080M} for planets
$b$ and $c$, which is expected considering our normalization procedure, we note that the photometry of HR8799 $e$
is also fully consistent with  \citet{2010Natur.468.1080M} measurements.
HR 8799$d$ results instead about 0.6 mag fainter in Ks band and 0.4 mag in H band. 
The discrepancy is however only marginally 
significant (about $2 \sigma$) as HR8799$d$ happens to be projected close to the AO outer 
working angle in our images (see below). Therefore photometry of planet $d$ is affected by larger errors (0.25 mag
in Ks band). Further observations are necessary to confirm the reality and physical nature of this variation.

\begin{table}
\caption{Photometry (absolute magnitudes) obtained from the H-band and the K$_{S}$-band data with the corresponding errors. Zero point photometric errors are not included. 
\label{tabphotometry}}
\begin{center}
\begin{tabular}{ccc }
\hline\hline
  Planet   &      H (mag)    &    Ks (mag)        \\ \hline
  b   &    $14.90\pm0.08$    &      $13.98\pm0.06$ \\
  c   &    $13.90\pm0.12$    &      $13.20\pm0.07$ \\
  d  &     $14.18\pm0.17$    &      $13.71\pm0.25$ \\
  e  &     $13.53\pm0.43$     &     $12.95\pm0.26$ \\   \hline
\end{tabular}
\end{center}
\end{table}

\subsection{Detection limits} \label{sub:detectionlimits}

\begin{figure}
\begin{center}
\includegraphics[width=8.0cm]{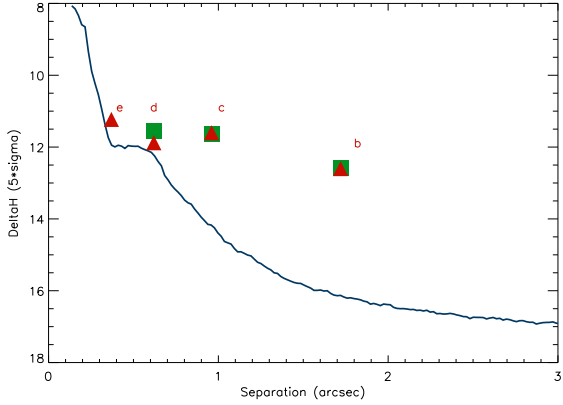}
\caption{5$\sigma$ contrast plot expressed in magnitude for the H-band data.
Overplotted our own measurement for the four HR8799 planets (red triangles) and the corresponding
measurements from \citet{2008Sci...322.1348M} as green squares. \label{hbandcontrast}}
\end{center}
\end{figure}

\begin{figure}
\begin{center}
\includegraphics[width=8.0cm]{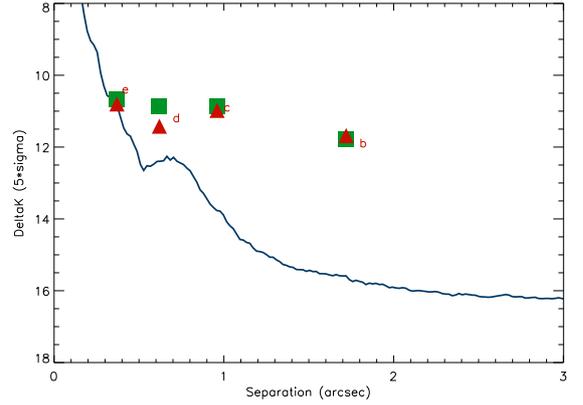}
\caption{5$\sigma$ contrast plot expressed in magnitude for the Ks-band data. 
Overplotted our own measurement for the four HR8799 planets (red triangles) and the corresponding
measurements from \citet{2008Sci...322.1348M} as green squares.
\label{kbandcontrast}}
\end{center}
\end{figure}

To calculate  the  detection  limit   as  a  function  of  the  angular separation from the  central star, 
we proceeded as  follows. Along a radial  direction,  we  calculated  the  standard  deviation  of  the
intensity over a box of $d$ $\times$ $d$  pixels (with $d$ equal to  the FWHM of
the  PSF i.e.  3 pixels  for FLAO  and PISCES)  and one  pixels step \citep{2005ApJ...625.1004M}. 
Values are then averaged  over the azimuthal  direction  and the sum of the contrast  $\Delta M$ at $5\sigma$ versus  
the angular separation is calculated normalizing with respect to the peak of the
planet suitably  re-scaled for the  correct flux and taking into account the flux losses due to the application of ADI. Fig.~\ref{hbandcontrast} and \ref{kbandcontrast}
show the  contrast obtained in H  and $K_{S}$ bands. The location of the
four  planets is  marked  with red  triangles.  For comparison, the
position of  the planets as determined  by \citet{2008Sci...322.1348M} are
showed  using green squares.
In Fig.~\ref{hbandcontrast} and \ref{kbandcontrast} the noise peak due to the outer working angle of the AO system
(which is expected to be at 0.49 arcsec for H and 0.65 arcsec for $K_{S}$\footnote{Following
\cite{2010SPIE.7736E..79R} the outer working angle (OWA) is computed as 
\begin{equation}
OWA=\lambda/(2d)~~arcsec
\end{equation}
\noindent 
where $d$ is the effective inter-actuator distance 
considering the number of correcting modes and the actuator pitch projected 
to the primary mirror, which is 27 cm. Hence, 
$d = 27 \sqrt{672/n_{modes}}$.   
When $n_{modes}=400$ (which is the value in general used in FLAO), $d=35$~cm.
Hence we obtain OWA=0.49 arcsec (H band, $1.65~\mu m$) and 0.65 arcsec (K band, $2.2~\mu m$)
It is important to mention that the extension of this halo spans about 
$\pm$ 0.06 arcsec from the OWA values reported above \citep[Fig.~17 in][]{2010SPIE.7736E..79R}.
These values agree very well with the position of the secondary maxima in 
the contrast vs separation curves. }) 
is clearly seen.
Such a feature is not usually present in detection limits obtained
with typical AO systems for 6-10 m class telescopes that achieve a much lower Strehl ratio than 
that delivered by the LBT FLAO system and PISCES camera
with our observations \citep{2005ApJ...625.1004M,2007ApJ...670.1367L,2007ApJS..173..143B,2010A&A...509A..52C}.


\section{Physical properties of planets around HR8799}
\label{s:phys}

We compared the near-IR properties of all HR8799 planets with those of  other substellar structures in order to investigate
systematic differences due to e.g. mass and age. First of all we compared the H and Ks 
magnitude observed for HR8799 planets with those obtained by \cite{2010ApJ...710.1627L} 
for field brown dwarfs (BDs). We then collected from the literature data for others low mass companions 
with direct imaging observations in the same photometric bands 
(Table~\ref{tabexoplimage}).
All these objects, together with the planets of HR8799, may be plotted in color-magnitude diagrams built with H and Ks filters. 

\begin{table*}
\caption{Photometry of others small mass companions adopted by literature. All listed magnitude are absolute magnitude\label{tabexoplimage}.}
\begin{center}
\begin{tabular}{lccccccl}
\hline\hline
Companion            &  D     & Mass & Age &  J              &    H             &    K             &  Ref.  \\ 
                     &  pc    & $M_{J}$ & Myr &              &                  &                  &        \\
\hline
2M1207b              &   52.4 &   5   &  8       &  16.40 $\pm$ 0.20 & 14.49 $\pm$ 0.21 & 13.33 $\pm$ 0.11 & 1,2  \\
1RXJ1609.1-210524b   &  140.0 &   8   &  5       &  12.17 $\pm$ 0.12 & 11.14 $\pm$ 0.07 & 10.44 $\pm$ 0.18 & 3,4  \\
AB Picb              &   47.3 &  13   & 30       &  12.80 $\pm$ 0.10 & 11.31 $\pm$ 0.10 & 10.76 $\pm$ 0.08 &  5   \\
HD203030b            &   40.8 &  23   & 130-400  &  15.08 $\pm$ 0.55 & 13.80 $\pm$ 0.12 & 13.16 $\pm$ 0.10 &  6   \\
HIP 78530b           &  156.7 &  23   &  5       &   8.94 $\pm$ 0.23 &  8.33 $\pm$ 0.22 &  8.14 $\pm$ 0.22 &  7   \\ 
CD 35 2722b          &   21.0 &  31   & 100      &  11.99 $\pm$ 0.18 & 11.14 $\pm$ 0.19 & 10.37 $\pm$ 0.16 &  8   \\
SR 12 AB c           &  125.0 &       &          &   9.9  $\pm$ 0.7  &  8.8  $\pm$ 0.9  &  8.6 $\pm$ 1.1   &  9   \\ 
HN Peg b             &   18.4 &  21   &  300     &  14.54 $\pm$ 0.06 & 14.08 $\pm$ 0.06 & 13.75$\pm$ 0.06  & 10   \\ 
Ross458 (AB) c       &  11.7  &  8.5  &  475     &  16.42$\pm$ 0.67 &16.78 $\pm$ 0.69 & 16.50 $\pm$ 0.69   & 11   \\ 
GSC 06214 -00210 b   & 145.0  &  17   &  11      &  10.5             &  9.6             &  9.1             & 12   \\ 
2M 044144 b          & 140.0  &  7.5  &  1       &                   &  9.89$\pm$0.10  &  9.21 $\pm$ 0.10  & 13   \\ 
CFBDSIR J1458+1013b  &  23.1  &  6.5  &  3000    &  19.84 $\pm$0.40  & 20.69 $\pm$ 0.27 & 21.02 $\pm$ 0.37 & 14   \\  
DH Tau b             & 143.5  &  11   &   1      &    9.65$\pm$0.05  & 9.03$\pm$0.04 &   8.33$\pm$0.02     & 15   \\ 

\hline
\end{tabular}
\tablebib{
 References: (1): \cite{2004A&A...425L..29C}; (2): \cite{2007ApJ...657.1064M} (3): \cite{2008ApJ...689L.153L}; 
             (4): \cite{2011ApJ...726..113I}; (5): \cite{2005A&A...438L..29C}; (6): \cite{2006ApJ...651.1166M}; 
             (7): \cite{2011ApJ...730...42L}; (8): \cite{2011ApJ...729..139W}; (9) \cite{2011AJ....141..119K}
            (10): \cite{2007ApJ...654..570L};  (11): \cite{2011MNRAS.414.3590B}; (12): \cite{2011ApJ...726..113I}; 
            (13): \cite{2010ApJ...714L..84T}; (14): \cite{2011ApJ...740..108L}; (15): \cite{2005ApJ...620..984I} }
\end{center}
\end{table*}

\begin{figure}
\begin{center}
\includegraphics[width=9.5cm,angle=0.0]{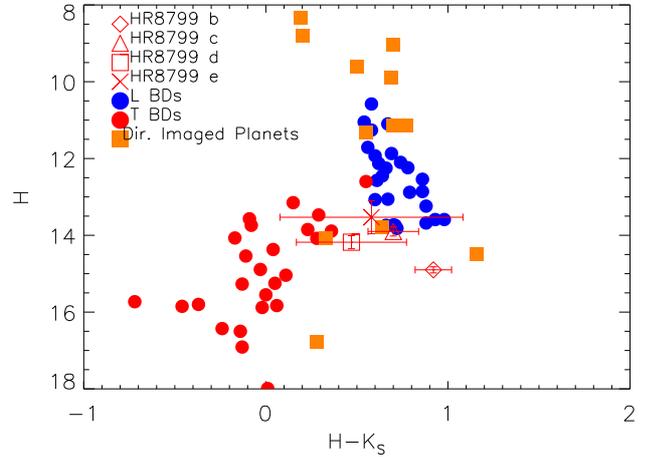}
\caption{H/H$-$K color magnitude diagram comparing the observed magnitudes of HR8799 planets with other cool objects 
in the field and known low mass companion. 
Three planets of the HR8799 system (c,d,e) are quite consistent  with the 
observed L and T spectral type BDs. HR8799 b is manifestly redder and fainter than L and T spectral type BDs separation. 
\label{hmagHKcolor}}
\end{center}
\end{figure}

The color-magnitude diagrams is shown in Figure~\ref{hmagHKcolor}. 
Three planets of  HR8799 ($cde$) have colors roughly consistent 
with L and T BDs sequence. In particular HR8799$c$ and HR8799$d$ are slightly underluminous with reference to the rim 
between L and T boundary, with HR8799$d$ redder than the colors of T BDs.
We also recall that our Ks photometry of $d$ has rather large error and is
0.6 fainter than \cite{2008Sci...322.1348M} one (see Sect.~\ref{s:results}). \cite{2008Sci...322.1348M} photometry
would imply a redder color, and a position slightly outside the field BD sequence.
HR8799$b$ has redder colors with respect the others HR8799 planets, and it is manifestly under 
luminous with respect not only the  L-T sequence of BDs but also with respect to the other planets 
of the HR8799 system (regardless of the systematic uncertainties in the photometry). 
A similar but even more extreme anomaly is that of \object{2M1207b}. 
To explain these locations in the color-magnitude phase space various ad-hoc
hypothesis were proposed, such as an occulting edge-on circumplanetary disk \citep{2007ApJ...657.1064M} 
or a collision afterglow \citep{2007ApJ...668L.175M}. The similar discrepancies occurring for 
HR8799$b$ and 2M1207b \citep{2011ApJ...735L..39B} and specific inconsistencies 
\citep[see][for details]{2011ApJ...732..107S} argue against such ad-hoc explanations 
suggesting a more general feature linked to cloud properties in low-gravity 
atmospheres.

In summary, we can confirm the faintness and redder NIR colors of HR8799b.
When considering only H/H$-$K  color-magnitude diagram, the positions of HR8799$c$,$d$, and $e$ 
is close to the field brown dwarf sequence, especially for the brightest planet $e$. 
Including information from other wavelengths clearly suggests the HR8799$c,d,$ and $c$ show significant
differences.
This is likely due to the presence of thick cloud layers and non equilibrium chemistry in 
their atmospheres \citep{2011ApJ...729..128C,2011ApJ...733...65B,skemer12}.
A more extensive discussion is given in the companion paper \citep{skemer12}.

\section{The architecture of the HR8799 planetary system}
\label{s:archit}

\subsection{Orbital fit}
\label{s:orbfit} 

Since the planet discovery papers \citep{2008Sci...322.1348M,2010Natur.468.1080M}, it resulted that 
all the planets in the HR8799 system orbit the star in the same direction and their orbits are roughly 
compatible with nearly pole-on circular orbits.
The cumulation of additional observational data \citep{2011A&A...528A.134B} and 
the detection of the some of planets in older images, thanks to a
reprocessing of the data \citep{2011ApJ...741...55S,2009ApJ...696L...1F,2009ApJ...694L.148L} allowed to further 
extend the time baseline of the observations.

The time coverage from 1998 to present epoch represents only a minor fraction of the orbital periods.
Nevertheless some indications of the actual orbits is emerging from the accumulated data.
The orbits of planets $b$ and $c$ are compatible with circular orbits seen nearly but not exactly pole-on.
Instead the orbit of planet $d$ is eccentric and/or seen at a different inclination \citep{2011A&A...528A.134B}.

A recent study of the configuration of the outer part of the HR 8799 system was presented by \cite{2011ApJ...741...55S},
taking advantage of the detection of the three outer planets in HST images taken in 1998.
They consider as the most likely solution a coplanar system in 4:2:1 mean motion resonance, with HR8799$d$
in a slightly eccentric orbit ($e=0.1$), as suggested in the dynamical stability analysis by \cite{2010ApJ...710.1408F}.

The increase of the time baseline of 1 yr is not decisive for a major revision of these results.
However, \cite{2011ApJ...741...55S} consider only the outer three planets and they restricted
to coplanar orbits. Furthermore, evaluation of dynamical stability of the proposed configuration was
not performed.

\begin{figure*}
\begin{center}
\includegraphics[width=4.5cm]{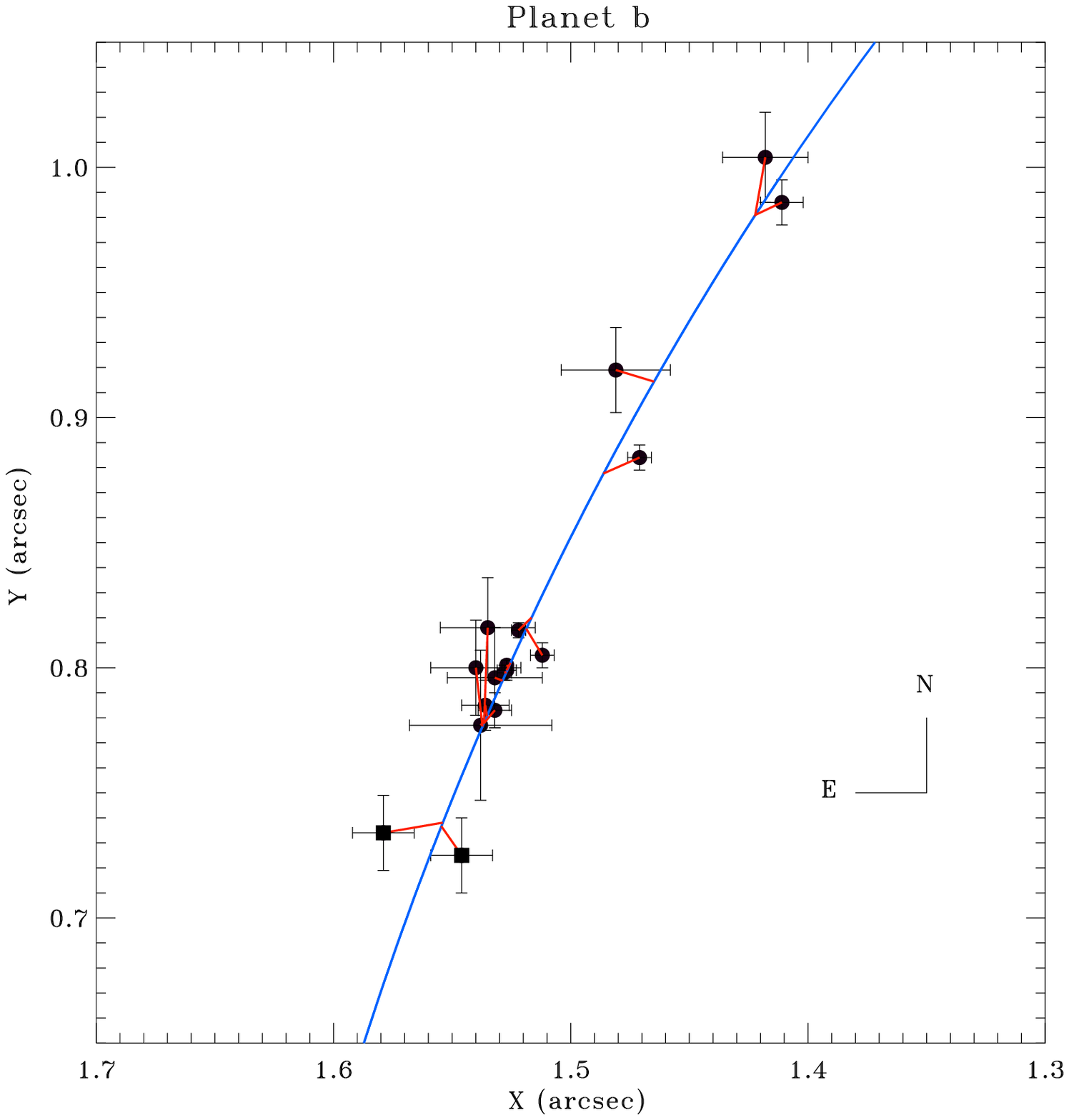}
\includegraphics[width=4.5cm]{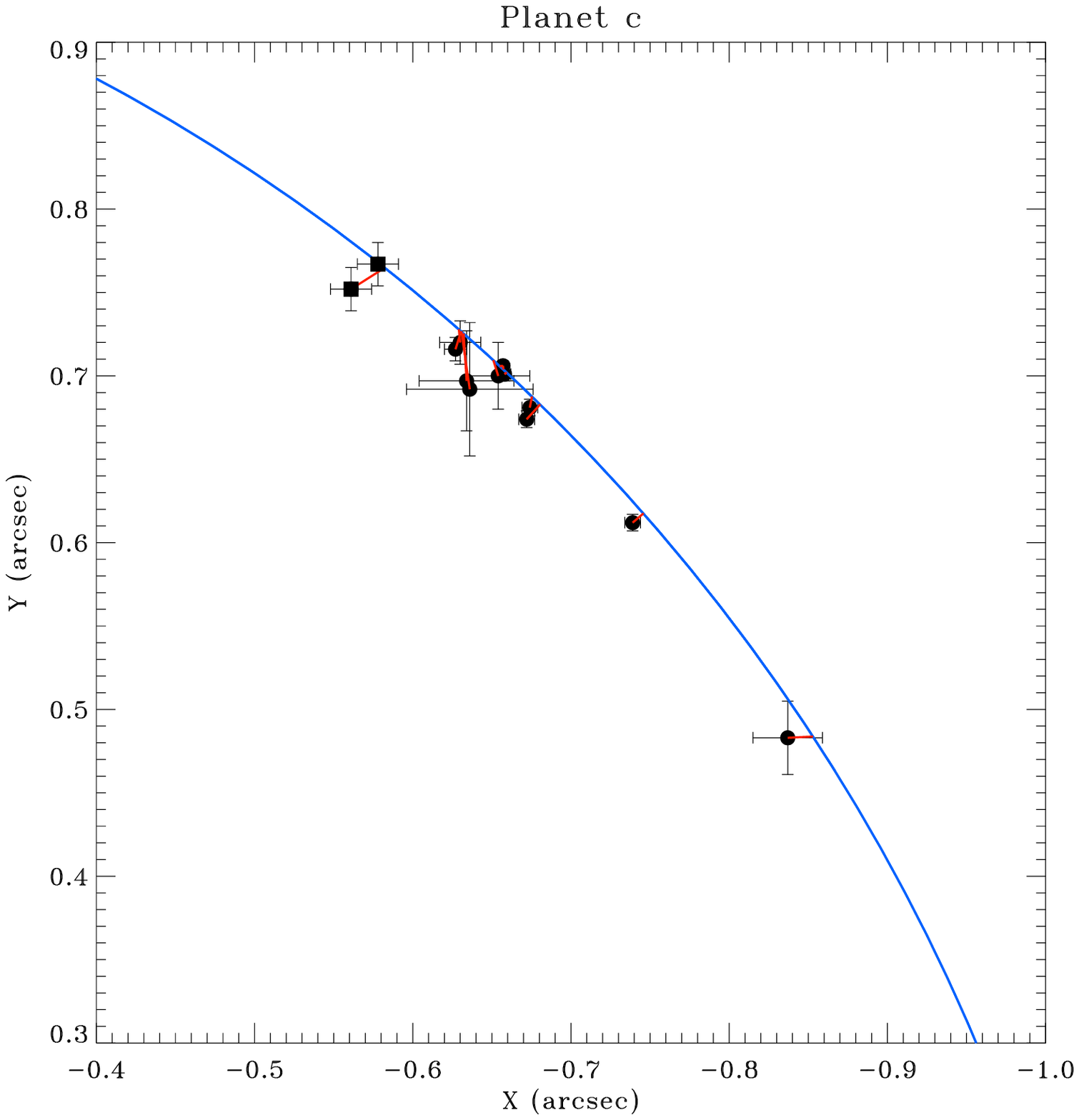}
\includegraphics[width=4.5cm]{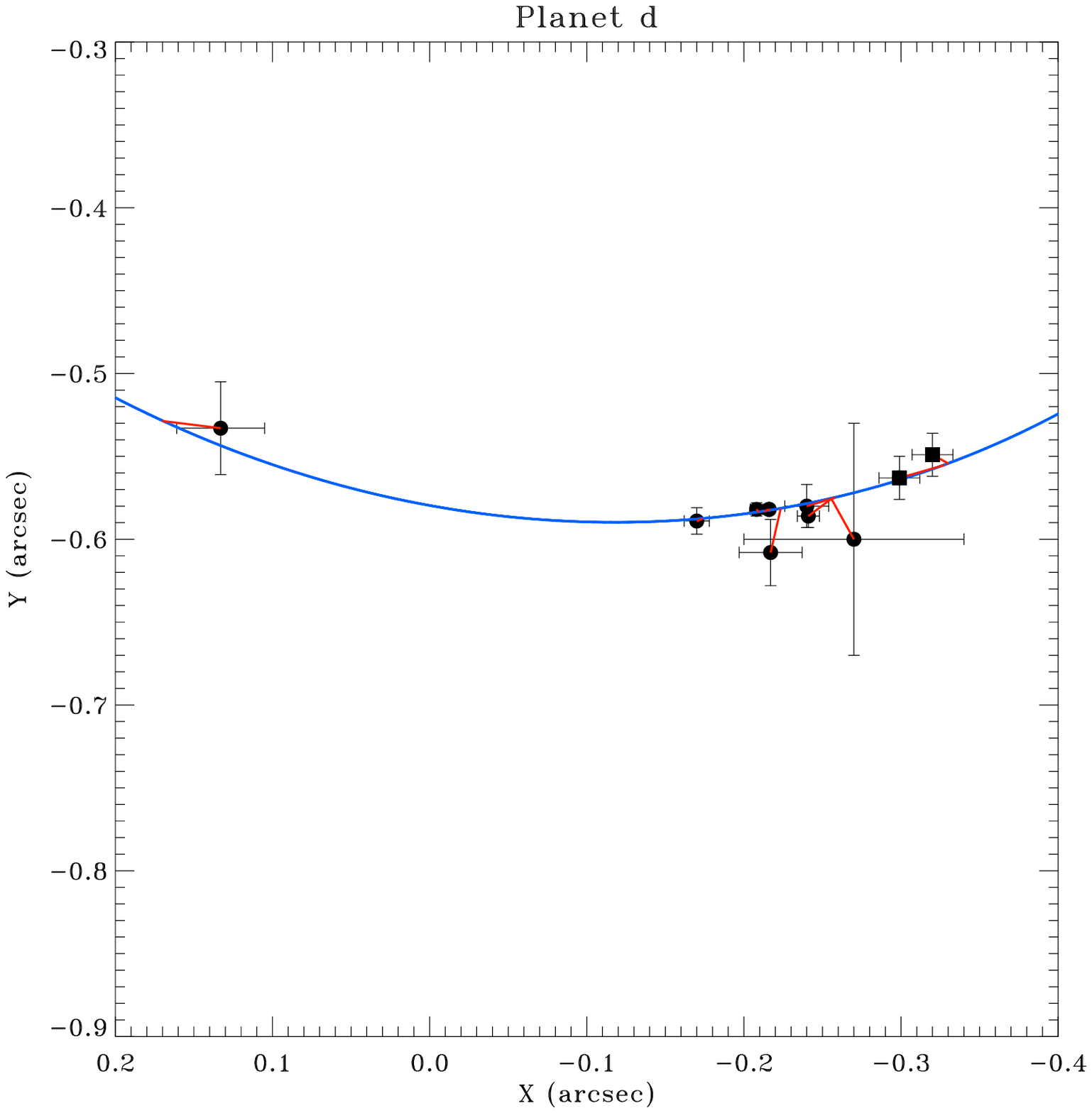}
\includegraphics[width=4.5cm]{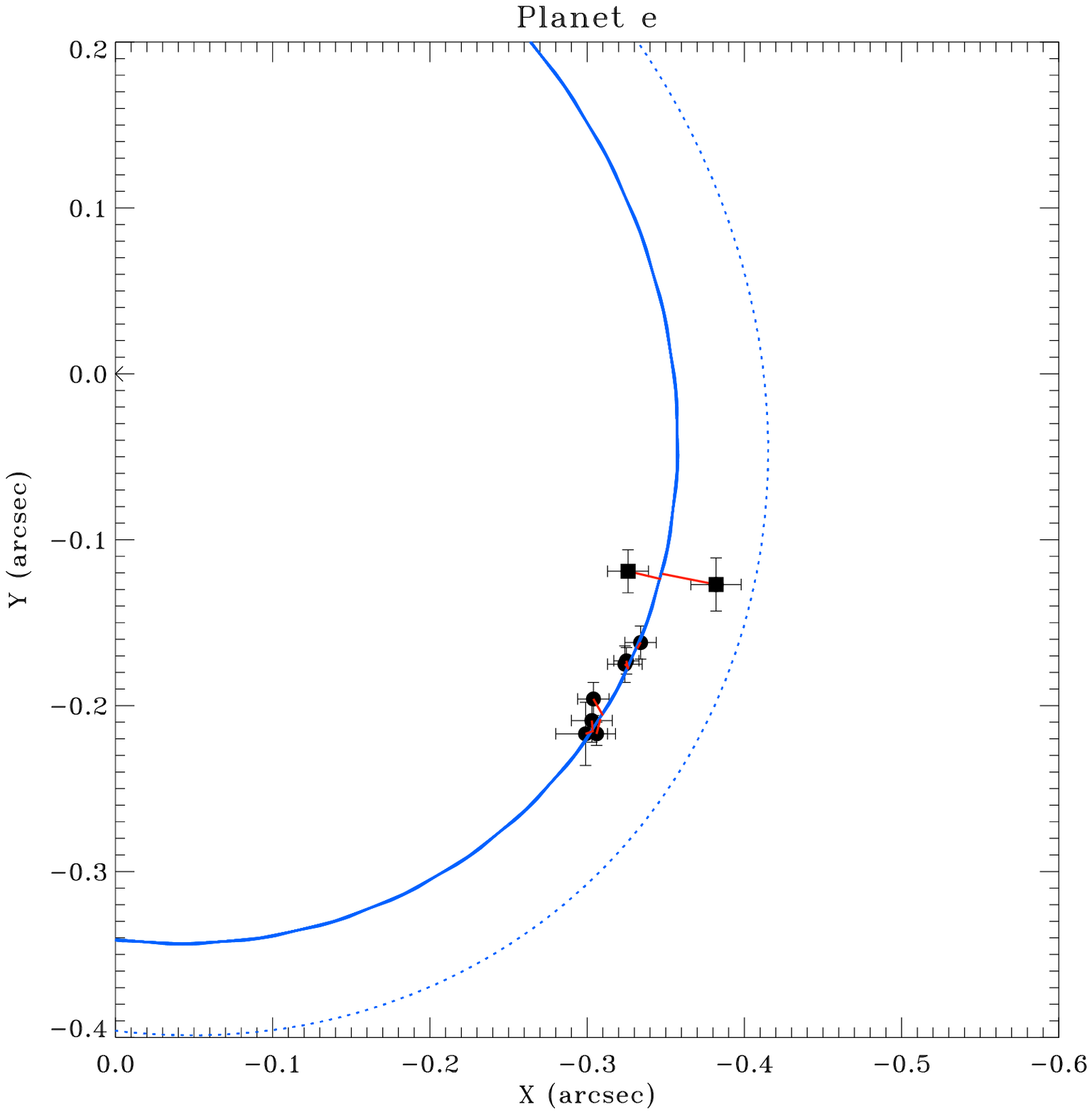}
\caption{Relative astrometry of HR 8799 planets. Overline are the \cite{2011ApJ...741...55S}
orbital solution for the outer planets and a coplanar circular orbital solution in 5:2 resonance for the inner planet  (blue lines, Case A
in Table \ref{t:orbit}).
For HR8799$e$, the dotted line shows the circular and coplanar orbital solution for 2:1 mean motion resonance with planet $d$, which
does not fit the observations.
Filled circles: literature results. Filled squares: data from the present paper.
Red lines connect the predicted and observed position for all the data points. }
\label{f:astrometry}
\end{center}
\end{figure*}

Fig.~\ref{f:astrometry} shows the relative astrometry of HR 8799 planets, as
compiled by \cite{2011A&A...528A.134B} plus the inclusion of data from \cite{2011ApJ...739L..41G}, 
\cite{2011ApJ...741...55S} and our own measurements (Tables~\ref{tabastrometryh} and \ref{tabastrometryk}).
Overplotted the \cite{2011ApJ...741...55S} orbital solutions for the three outer planets and 
our own solution for the inner planet (see below for the fitting procedure).

\begin{figure*}
\begin{center}
\includegraphics[width=4.5cm]{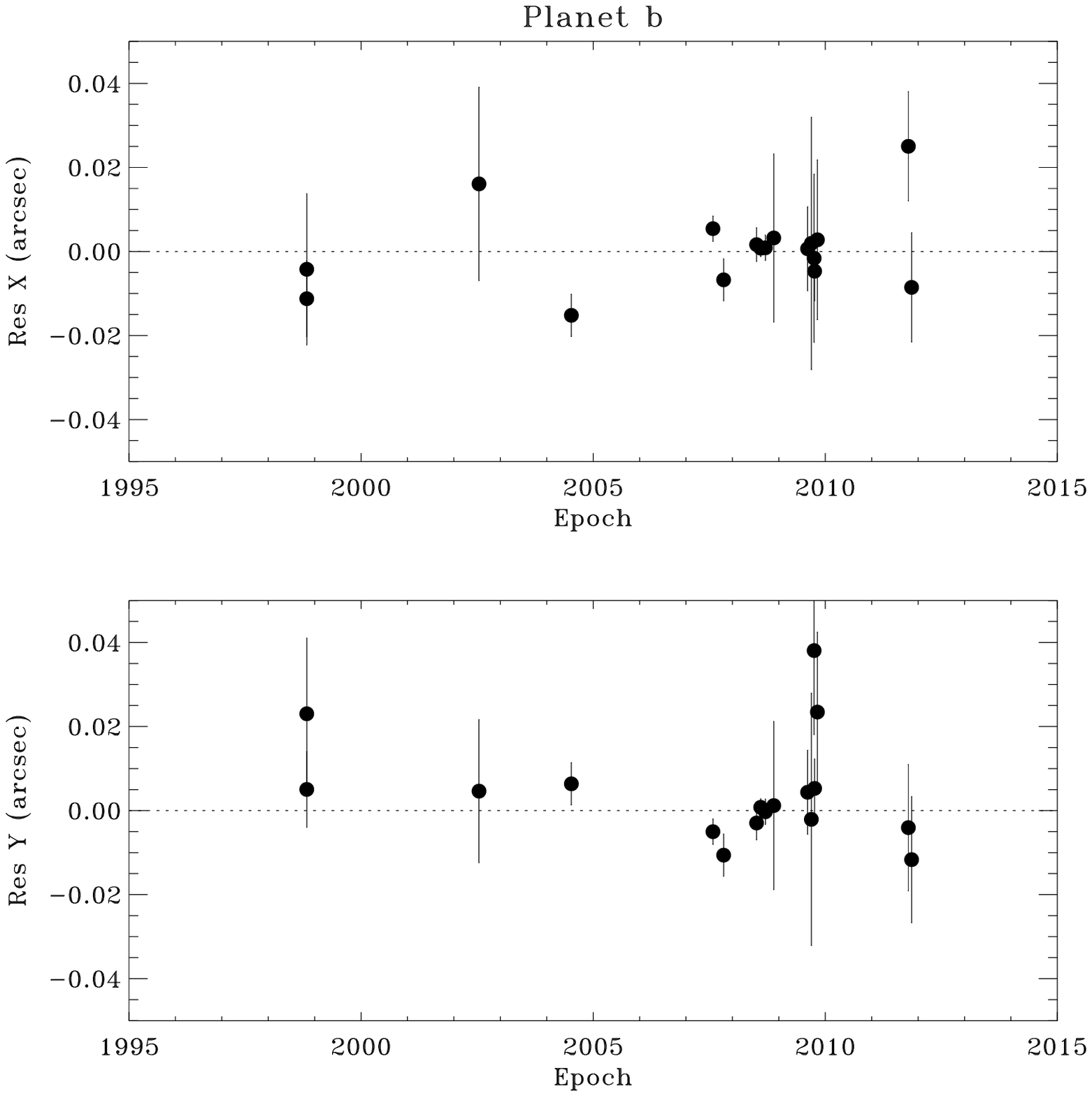}
\includegraphics[width=4.5cm]{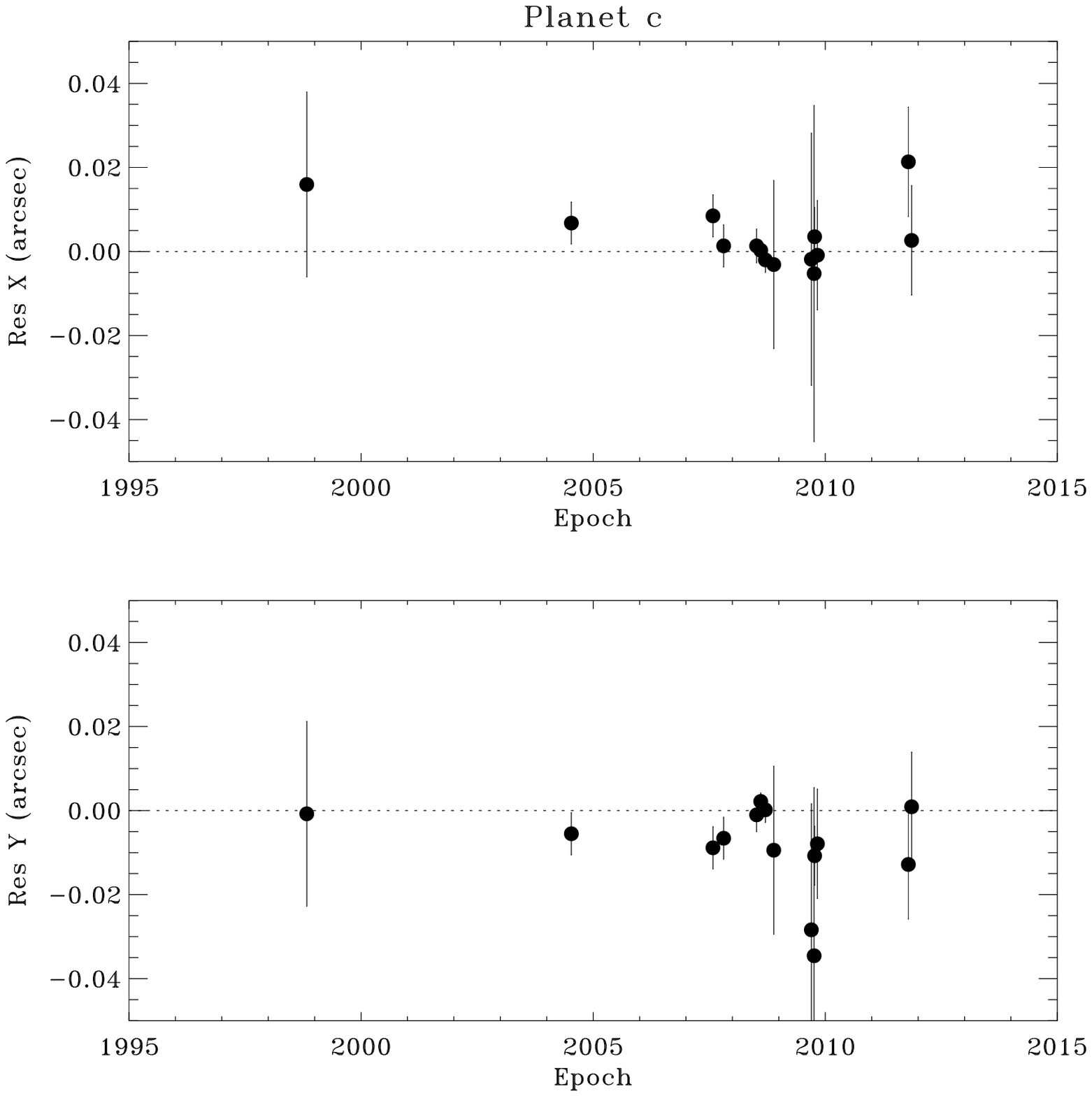}
\includegraphics[width=4.5cm]{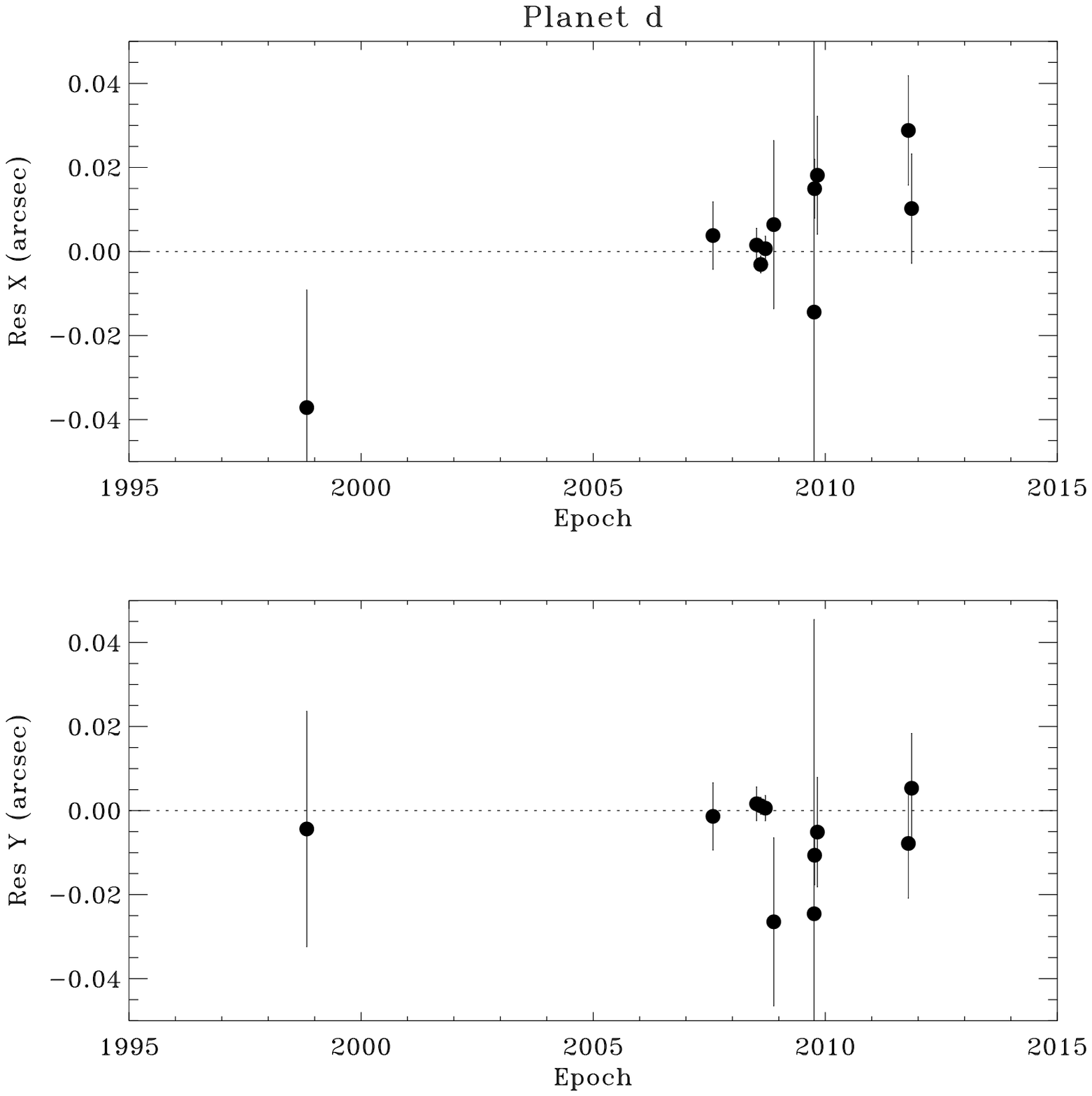}
\includegraphics[width=4.5cm]{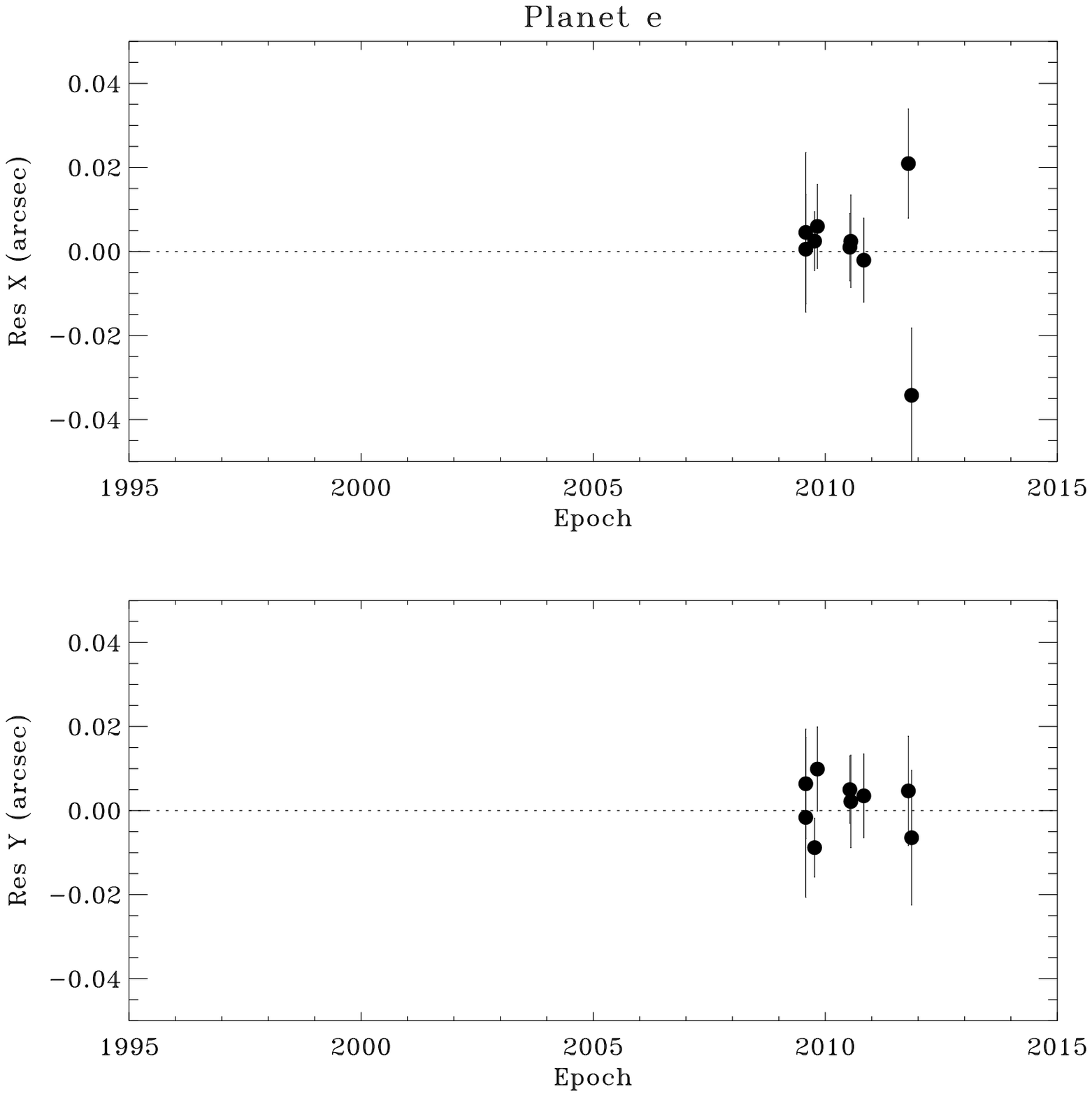}
\caption{Residuals in $x$ and $y$ coordinates vs time for each of the planets with respect to the orbital solution plotted
in Fig.~\ref{f:astrometry} (Case A in Table \ref{t:orbit}).}
\label{f:residuals}
\end{center}
\end{figure*}

New data are consistent with the previous works and follow the best-fit orbit 
derived in \cite{2011ApJ...741...55S} for the three outer planets.
If we adopt the \cite{2011ApJ...741...55S} orbital solution, planet $e$ can not be in a circular and coplanar
orbit with a 2:1 mean motion resonance with $d$. Instead, a 5:2 resonance represents
a satisfactory fit to the data, when assuming circular orbit and coplanarity with the
other planets (Fig.~\ref{f:astrometry}, Case A in Table~\ref{t:orbit}).

\begin{figure*}
\begin{center}
\includegraphics{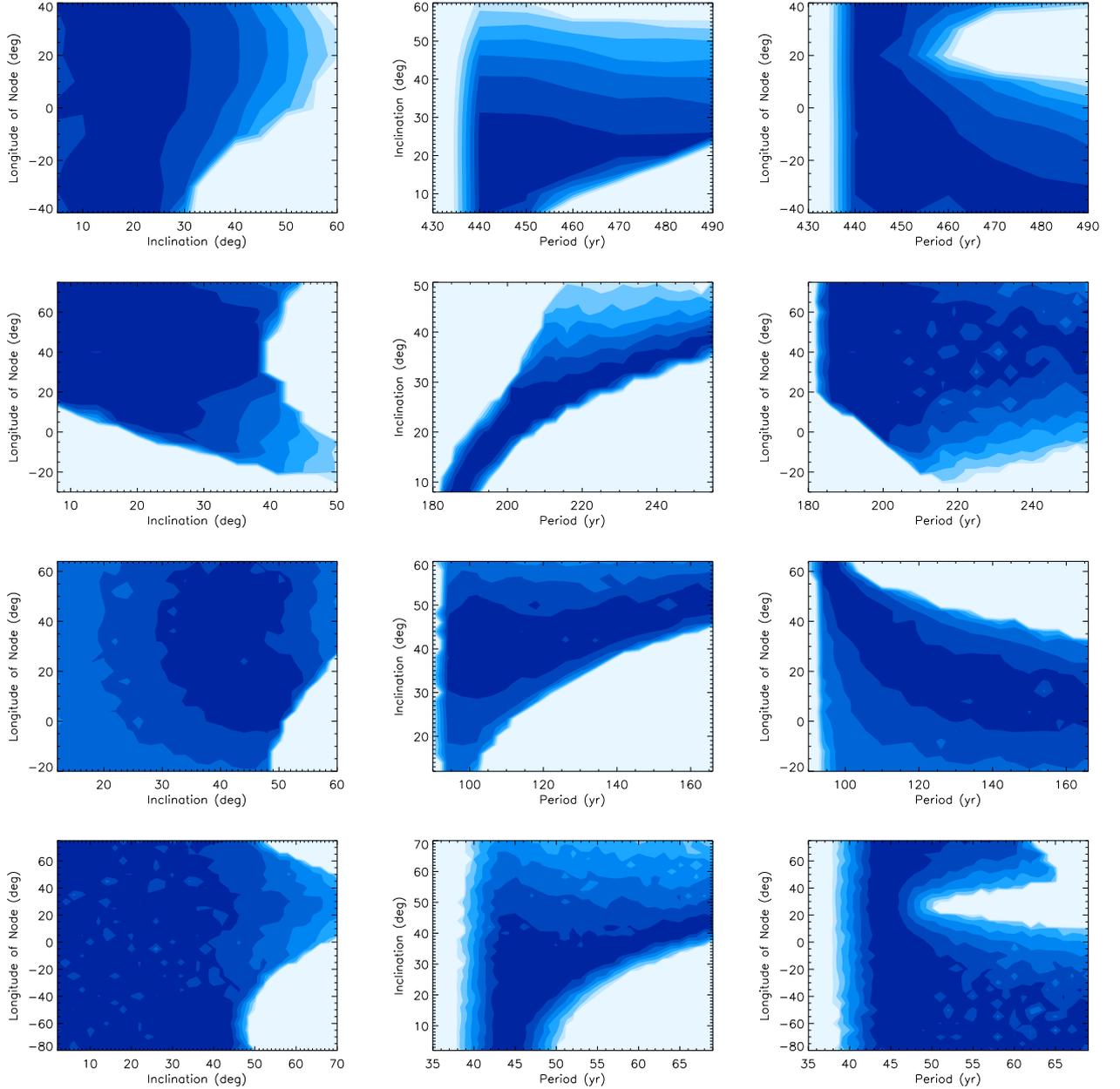}
\caption{Reduced $\chi^2$ surfaces for orbital period, inclination and longitude
of node for HR8799 planets assuming circular orbits. Each plot row refers to one of the planets, starting from
$b$ (top) to $e$ (down). The plotted  levels refer to reduced $\chi$ values in step of 0.20 starting from 2.0, with
darker area correspinding to lower $\chi$ values.}
\label{f:chi2}
\end{center}
\end{figure*}

This orbital solution represents only one of the possible ones fitting the observational data.
Unique determination of orbital parameters is not yet possible as the observations
cover only a minor fraction of the orbital periods.
To somewhat complement the \cite{2011ApJ...741...55S} study, that assumes coplanarity between
the three outer planets, we focused our attention here on non-coplanar configurations.
We restrict our analysis to circular orbits to reduce the number of parameters.
However, this approximation is likely not realistic considering that secular perturbations among planets 
causes some eccentricity pumping, with average
values of about 0.03-0.05. We also considered only a stellar mass of $1.56~M_{\odot}$, the preferred
value in \cite{2011ApJ...741...55S}.
Our analysis of possible orbital solutions was performed in two steps:

\begin{itemize}
\item
MonteCarlo simulations for a broad exploration of the orbital solutions compatible with
the data and correlations between the orbital parameters (Fig.~\ref{f:chi2}).
This allows us to identify the possible ranges to bound the least-square orbital fitting 
and to identify appropriate initial guesses.
\item
Best-fit least-square orbital solution for the four
planets simultaneously using the Levenberg-Marquard minimization algorithm as
implemented in the IDL routine MPFIT.
The program allows to fix some of the parameters (e.g. impose null eccentricity) or to
tie some parameters of one planet to those of another planet (e.g for coplanar orbits 
or for imposing orbital resonances). 
\end{itemize}

When forcing all consecutive planets to be involved in 2:1 resonance (8:4:2:1 resonance involving 
all the known planets) we found orbital solutions that nicely fit the available data.
The formal best-fit case is listed as case B in Table ~\ref{t:orbit}.
A notable property of this orbit is the strong deviation from coplanarity involving the
inner planet $e$, larger than $50^{\circ}$ with respect to planet $d$ (when including both 
$i$ and $\Omega$ into account).
Forcing planets $bcd$ to be in a 4:2:1 resonance, planets $de$ to be in a 5:2 resonance, and 
assuming all orbits are circular yields
the orbital parameters labeled as case C in Table ~\ref{t:orbit}.
In this case the deviations from coplanarity are or the order of $10-20^{\circ}$ and can become
much smaller when allowing for some eccentricity of the orbits (see Case A).
Such differences in the orientation of HR8799$e$ orbit depending on the chosen
resonance with $d$ are independent on the orbit of planet $b$. We found a similar
behaviour when adopting an outer orbit for $b$, outside 2:1 resonance with $c$ 
following \cite{2012arXiv1201.0561S}.

Additional constraints on the orientation of the components of the HR 8799 system
are available for the central star, for which \cite{2011ApJ...728L..20W} derived
$i \ge 40^\circ$ from asteroseismology, and for the outer debris disk, for which
\cite{2009ApJ...705..314S} found $i \le 25^\circ$. The best-fit solutions
listed in Table \ref{t:orbit} show that the outer three planets have intermediate
values, with hints of a trend in inclination from $b$ to $d$ toward more pole-on 
orientations inside (cases B and C). 
The relative inclination of $e$ instead depends critically on the adopted
orbital period, as discussed above. 
The two orbital solutions (B and C) while fitting in the same way the current data
will diverge significantly within 2-3 years.

\begin{table}
\caption{Orbital solution considered in this paper. Case A: Orbital solution from \cite{2011ApJ...741...55S}
+ our fit for planet $e$ (circular orbit, coplanar with other planets and in 5:2 mean motion resonance with planet $d$). 
Case B: best fit orbital solution imposing circular orbits and 8:4:2:1 mean motion resonance. Case C: 
best fit orbital solution imposing circular orbits and 4:2:1 mean motion resonance for planets $bcd$ and 5:2 for $de$}
\label{t:orbit}
\begin{center}
\begin{tabular}{c c c c }
\hline\hline
  Parameter                &    A    &   B   &   C    \\ \hline

$P_{b}$ (yr)               &   449.7  &  448.44   &    448.44    \\
$i_{b}$  ($^{\circ}$)      &    28.0  &   20.07   &     20.07    \\  
$\Omega_{b}$ ($^{\circ}$)  &    35.5  &   22.52   &     22.53     \\  
$e_{b}$                    &     --   &     --    &    --   \\
$\omega_{b} $ ($^{\circ}$) &     --   &     --    &    --   \\
$T0_{b}$ (yr)              &  1997.55 &  2015.37  &  2015.37     \\
$a_{b}$ (AU)               &    68.08 &    67.94  &  67.94     \\ \hline

$P_{c}$ (yr)               &   224.9  &   224.21  &   224.22    \\
$i_{c}$  (deg)             &    28.0  &    27.78  &    27.78    \\  
$\Omega_{c}$ ($^{\circ}$)  &    35.5  &    45.86  &    45.86   \\  
$e_{c}$                    &     --   &      --   &    --   \\
$\omega_{c} $ ($^{\circ}$) &     --   &      --   &    --   \\
$T0_{c}$ (yr)              &  1844.10 &  1838.89  &   1838.89    \\
$a_{c}$ (AU)               &    42.89 &    42.80  &  42.80     \\ \hline

$P_{d}$ (yr)               &   112.4  &   112.11  &   112.11    \\
$i_{d}$  ($^{\circ}$)      &    28.0  &    40.43  &    40.43    \\  
$\Omega_{d}$ ($^{\circ}$)  &    35.5  &    36.76  &    36.76   \\  
$e_{d}$                    &    0.10  &      --   &    --   \\
$\omega_{d} $ ($^{\circ}$) &    80.2  &      --   &    --   \\
$T0_{d}$ (yr)              &  1992.31 &  1965.20  &   1965.20    \\
$a_{d}$ (AU)               &    27.01 &    26.97  &     26.97     \\ \hline

$P_{e}$ (yr)               &   44.96  &    56.05  &    44.84   \\
$i_{e}$  ($^{\circ}$)      &   28.0   &    31.13  &    20.09    \\  
$\Omega_{e}$ ($^{\circ}$)  &   35.5   &   -62.81  &    43.39   \\  
$e_{e}$                    &    --    &     --    &    --   \\
$\omega_{e} $ ($^{\circ}$) &    --    &     --    &    --   \\
$T0_{e}$ (yr)              & 1987.08  &   1996.70 &   1986.14    \\
$a_{e}$ (AU)               &   14.66  &     16.99 &    14.64     \\ \hline

$M_{star}$ ($M_{\odot}$)   &   1.56   &     1.56  &   1.56    \\

\end{tabular}
\end{center}
\end{table}

\begin{figure*}
\begin{center}
\includegraphics[width=4.5cm]{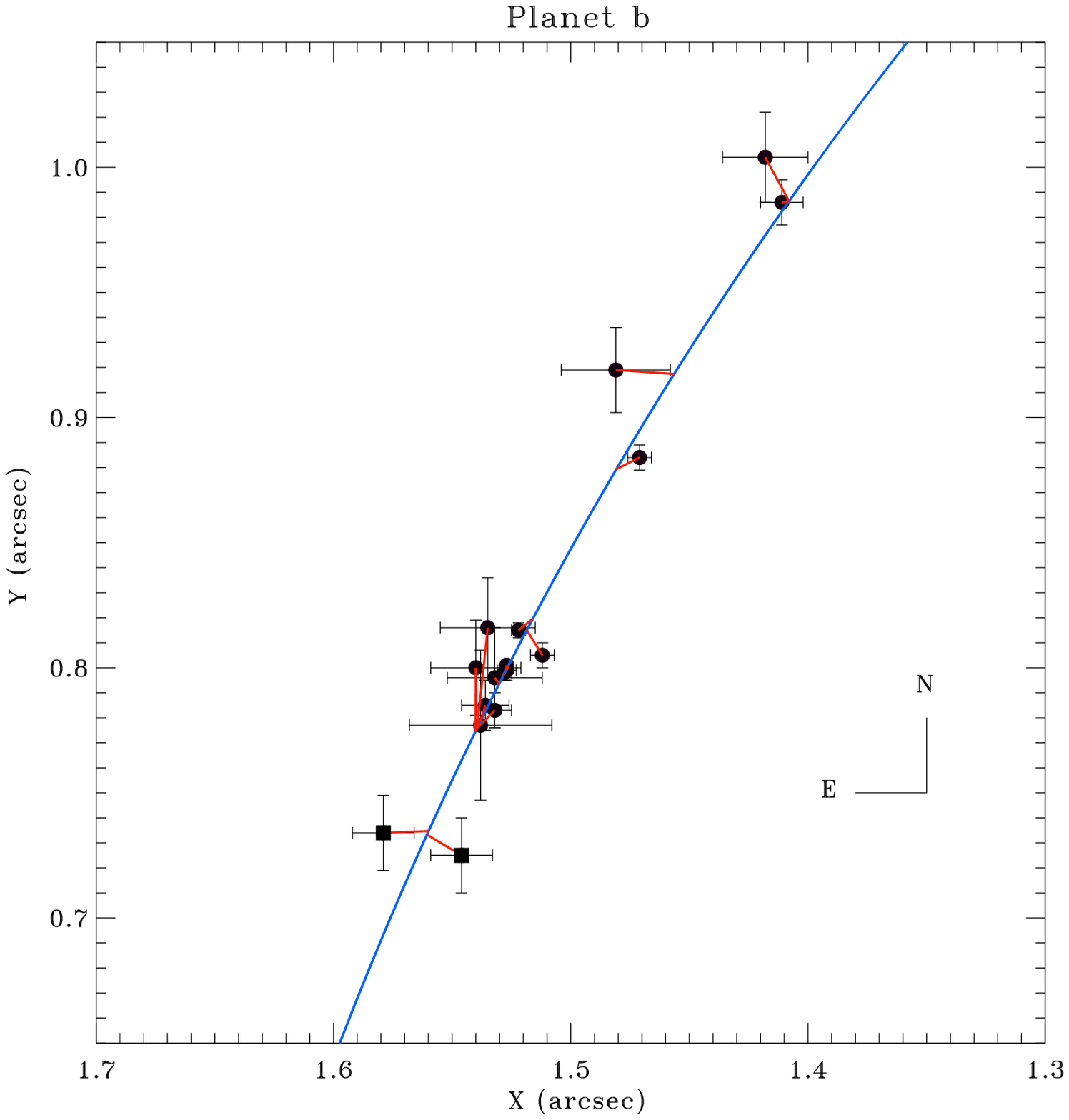}
\includegraphics[width=4.5cm]{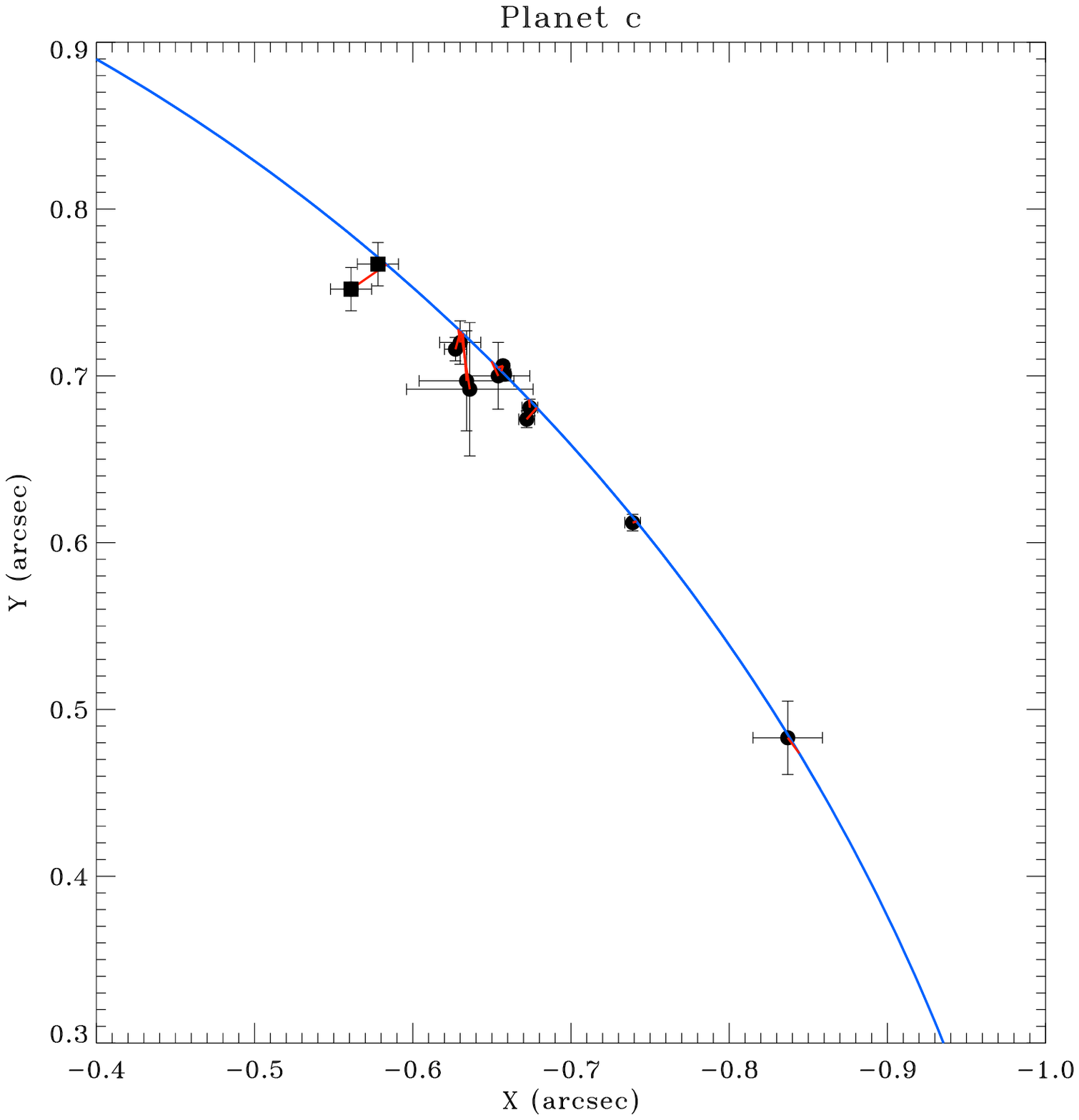}
\includegraphics[width=4.5cm]{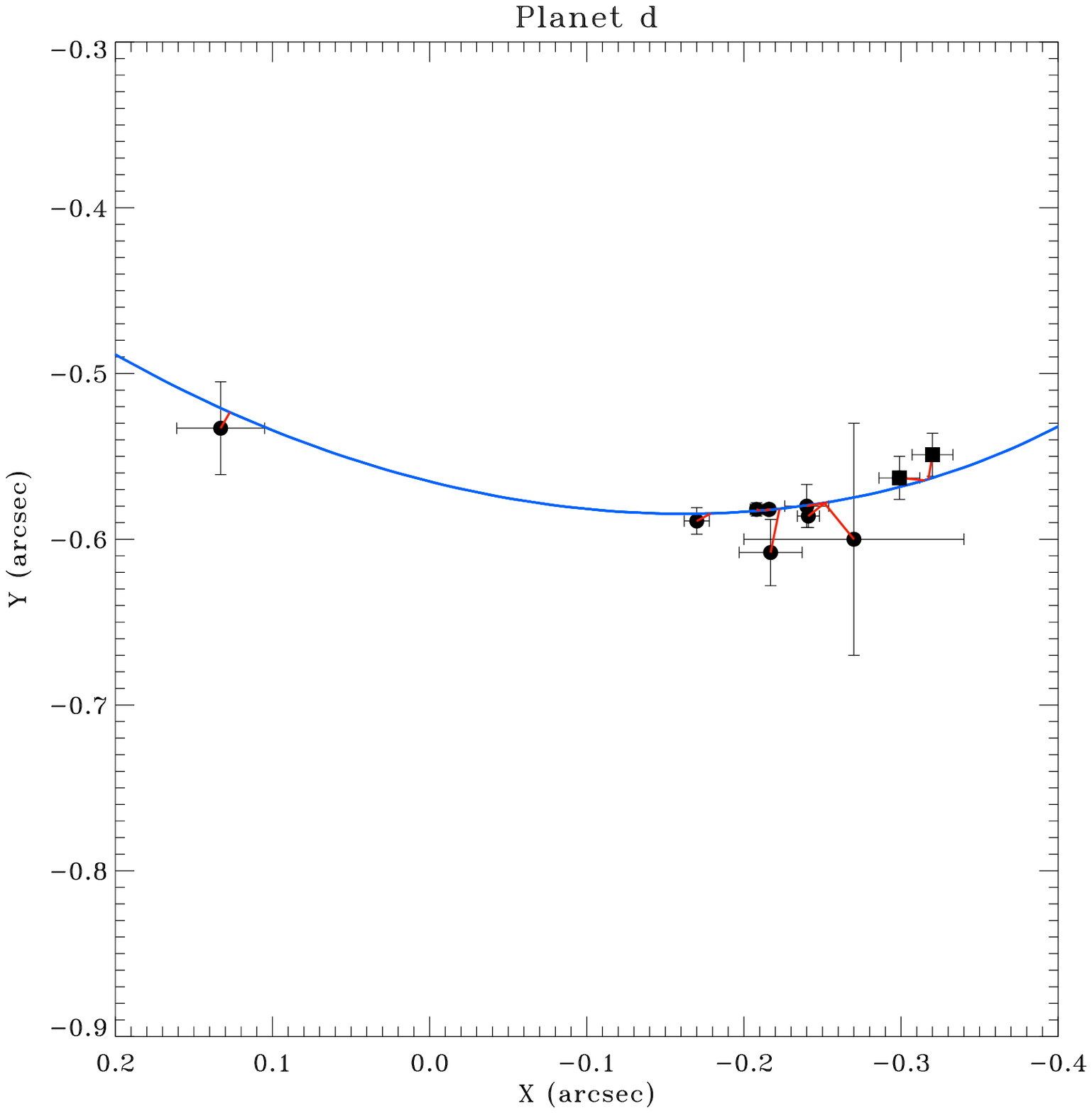}
\includegraphics[width=4.5cm]{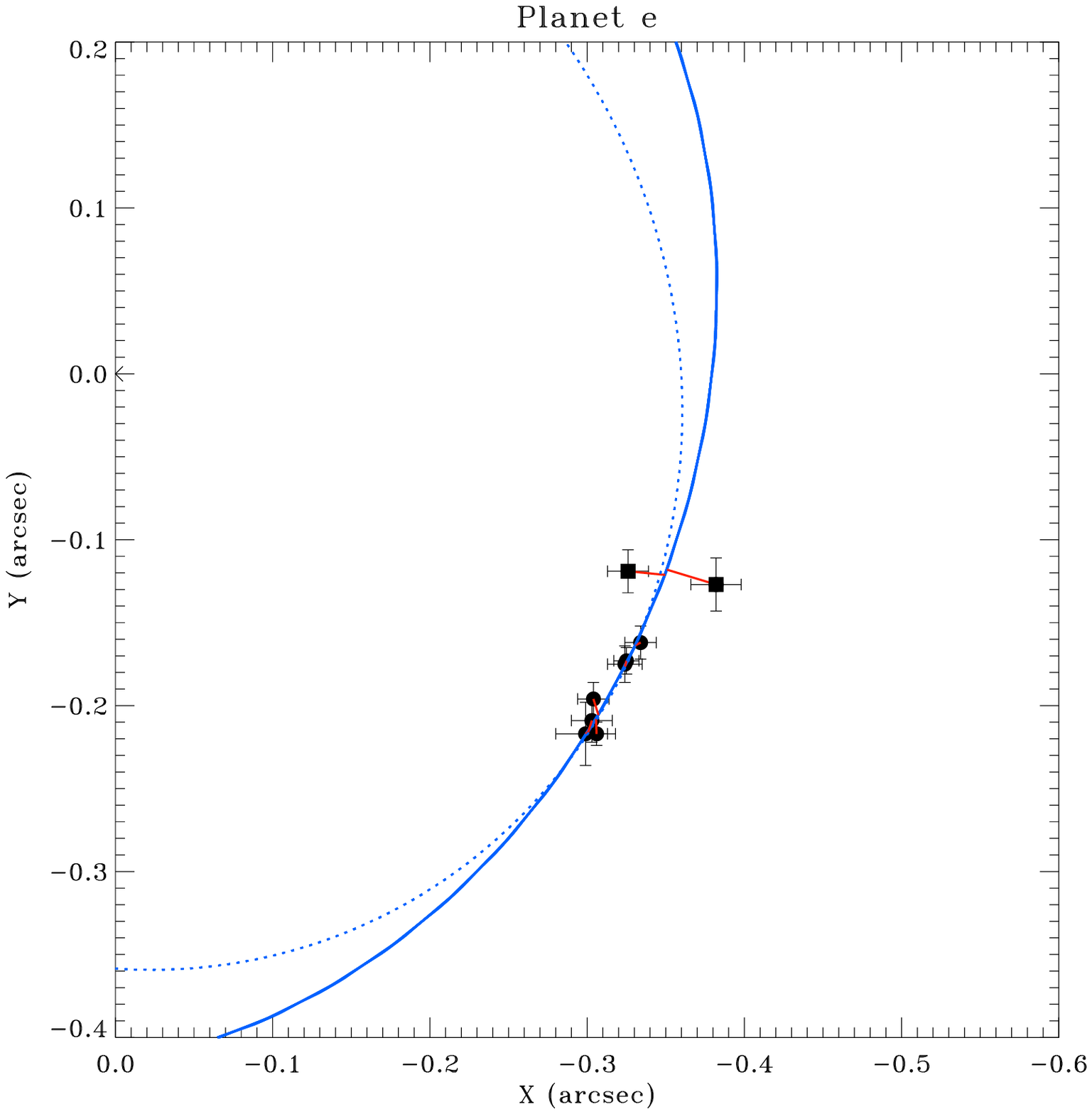}
\caption{Relative astrometry of HR 8799 planets available in literature. Blue lines show the adopted
orbital solution (Case B in Table \ref{t:orbit}, non coplanar circular orbits). Red lines connect the predicted and observed position for all the data points.
For HR8799$e$, the dotted line shows the orbital solution for 5:2 mean motion resonance with planet $d$ (Case C in Table \ref{t:orbit}).}
\label{f:astrometry1}
\end{center}
\end{figure*}

\begin{figure*}
 \begin{center}
\includegraphics[width=4.5cm]{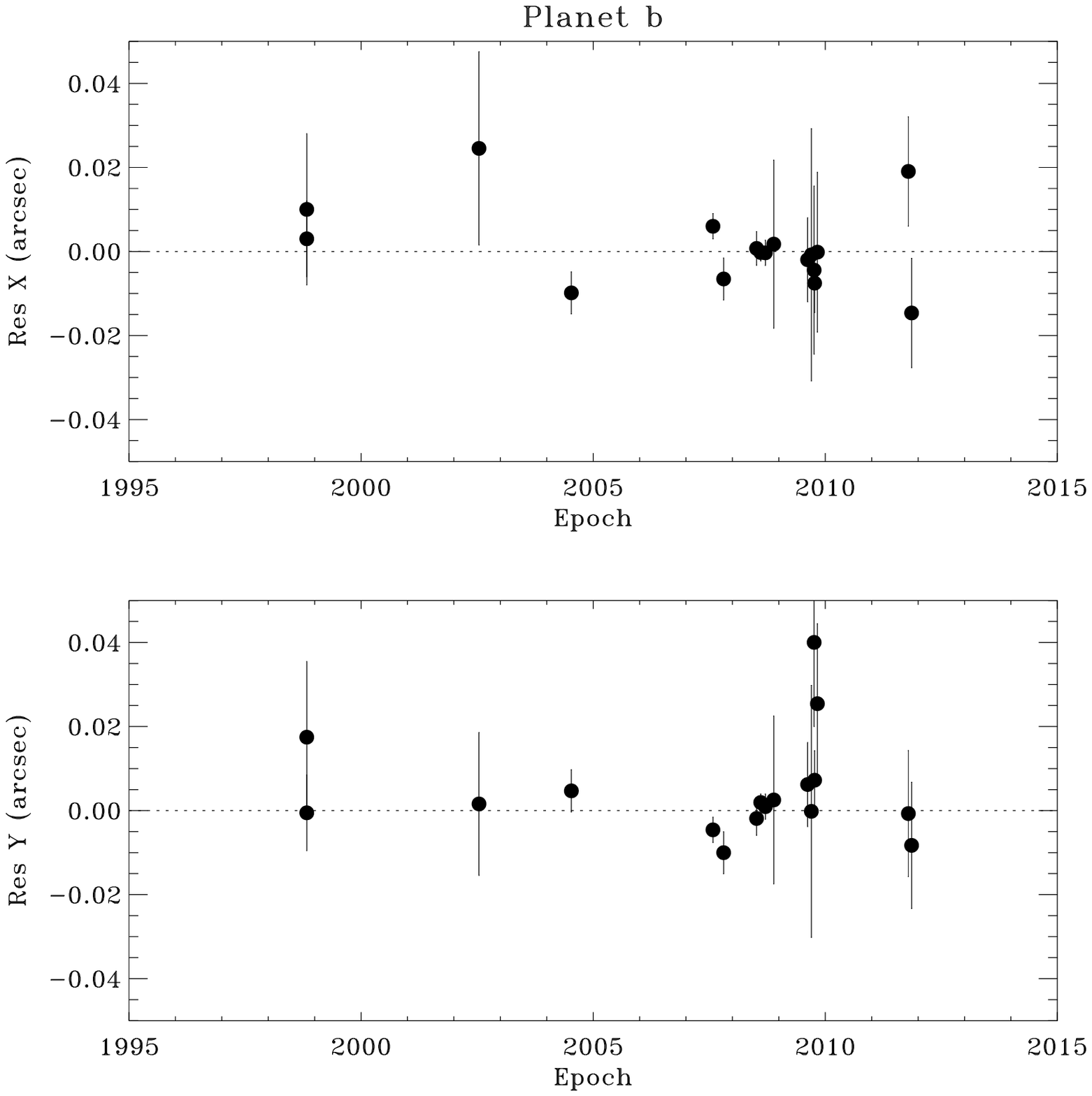}
\includegraphics[width=4.5cm]{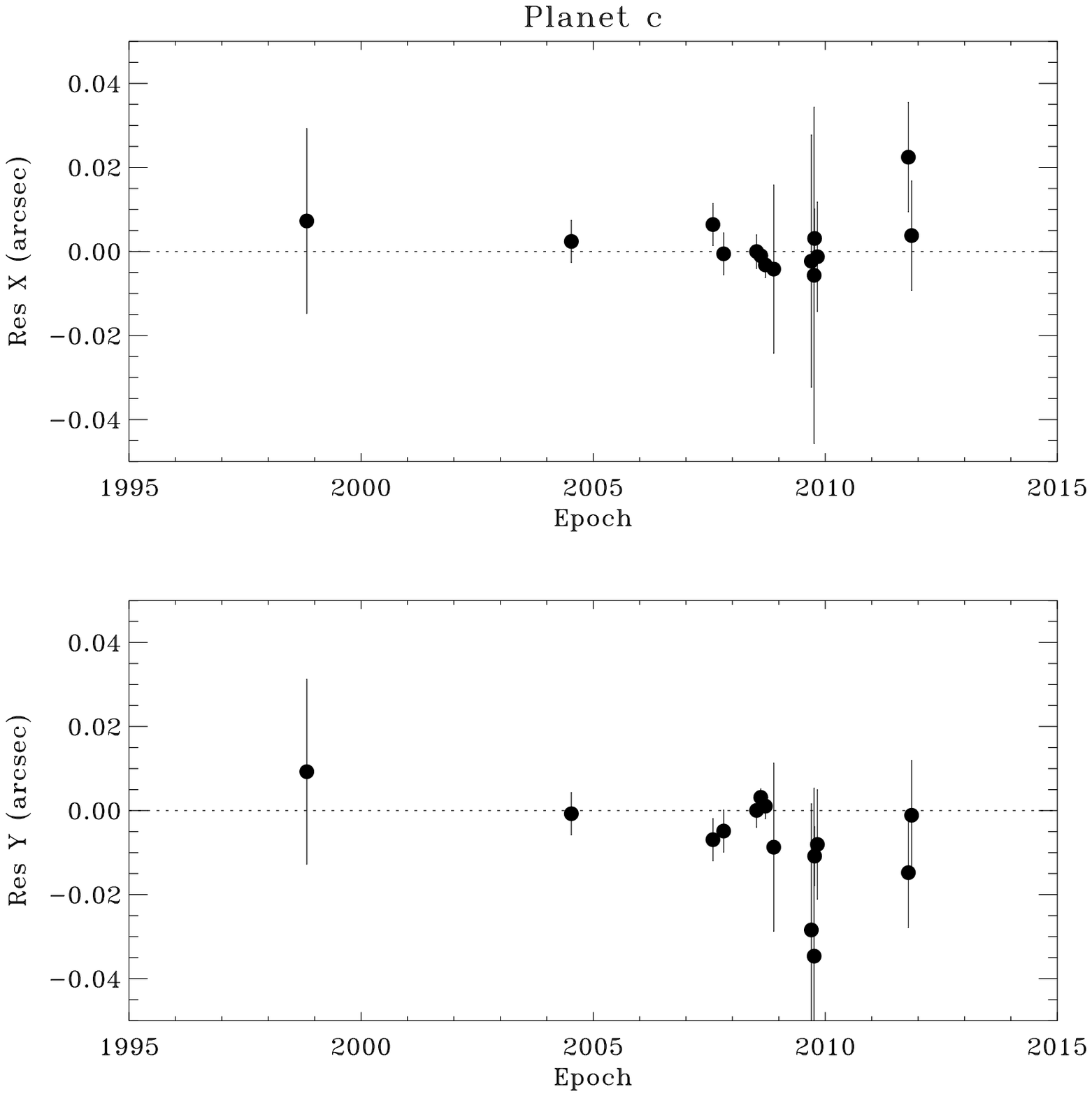}
\includegraphics[width=4.5cm]{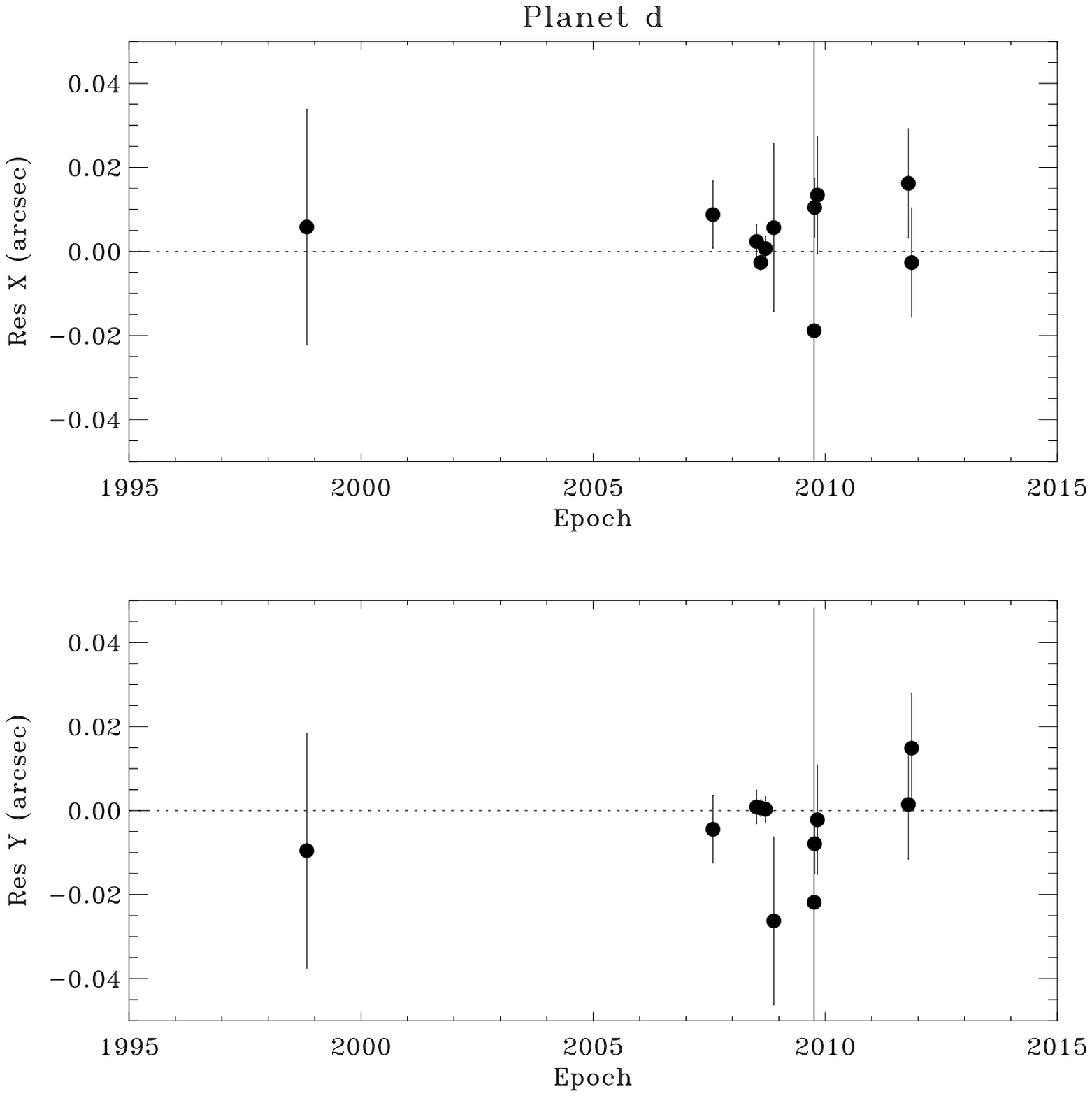}
\includegraphics[width=4.5cm]{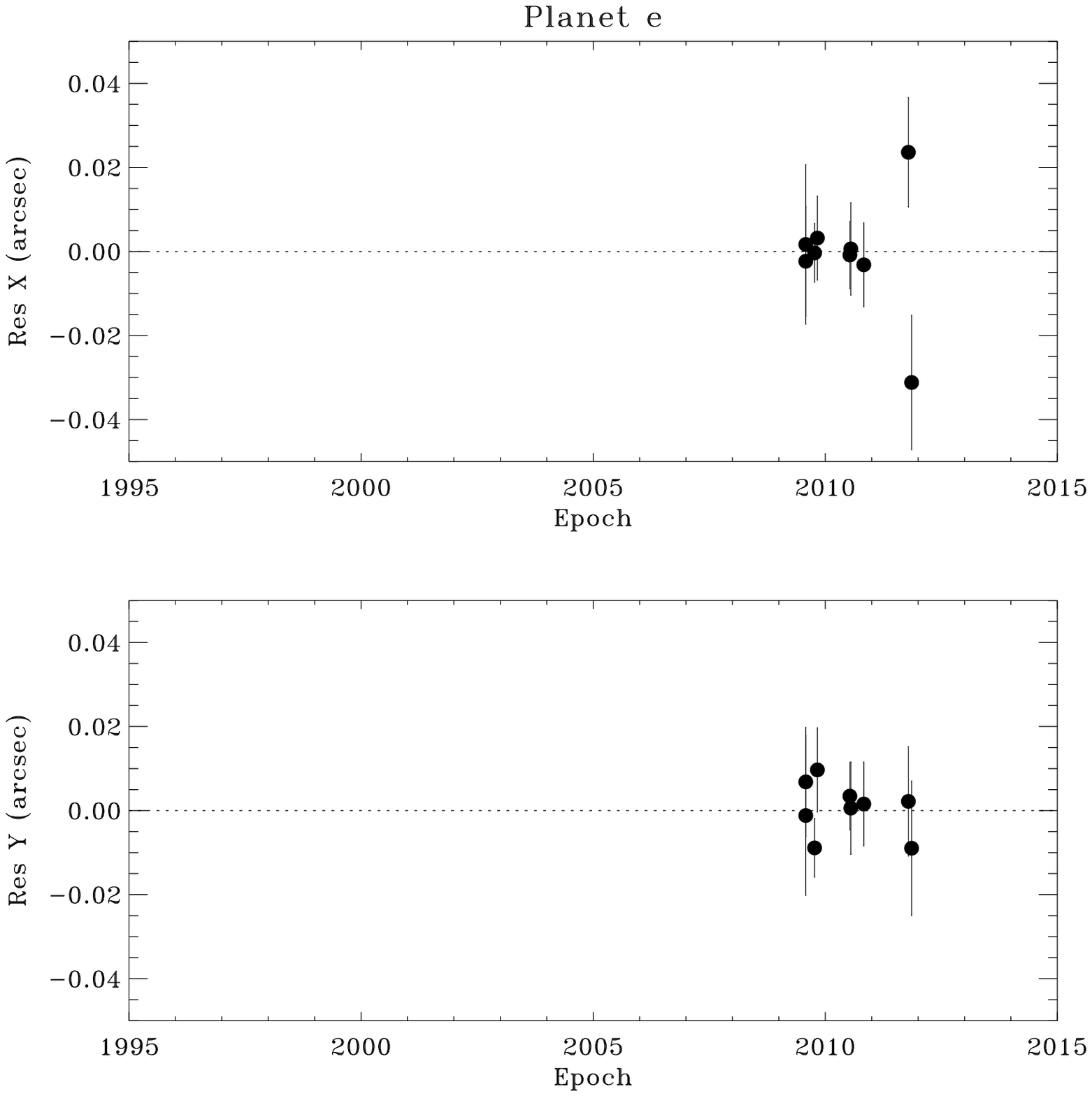}
\caption{Residuals in $x$ and $y$ coordinates vs time for each of the planets with respect to the orbital solution plotted
in Fig.~\ref{f:astrometry1} (Case B in Table \ref{t:orbit}).}
\label{f:res1}
\end{center}
\end{figure*}

\subsection{Dynamical stability}

After these explorations based only on observational data, we consider the additional
clues that can be derived from the dynamical stability of the system.

Previous dynamical studies of the system considering the three planet systems (without planet $e$), 
found that only a minor fraction of the orbital solutions compatible 
with the astrometric observations is dynamically stable.
Stability is favoured by the occurrence of a 2:1 mean motion orbital resonance 
between each couple of consecutive planets,
low eccentricity of their orbits and by planetary masses at the low end of the plausible values
derived from their luminosity and stellar age \citep{2009MNRAS.397L..16G,2009A&A...503..247R,2010ApJ...710.1408F}.

Recently, \cite{2012arXiv1201.0561S} presented a study of the dynamical stability of the 4-planet system.
They found that the system is strongly unstable with 
a few solutions that barely survive 
for ages comparable to the age of the system. Planet $e$ is marginally fit by these stable solutions, and 
\cite{2012arXiv1201.0561S} proposed that HR8799$e$ astrometry is systematically biased toward larger separations.
Additional, more limited, dynamical studies of the 4-planet system were performed in the discovery paper by 
\citet{2010Natur.468.1080M} and in \citet{2011ApJ...729..128C} with similar results.

We numerically integrated the orbits of the planets around the star 
in a full 5 body problem using the numerical integrator RADAU \citep{1985dcto.proc..185E}.
The timespan of the integration was fixed to 25 Myr and the computed orbital elements were used 
to estimate the stability properties using the Frequancy Map Analysis (FMA)
technique as in \cite{2005A&A...442..359M}. We studied the shift in frequency of the 
perihelion longitude of the middle planet as stability indicator. 
The phase space of a four planet system is wide so we devised the following strategy. 
We set the two middle planets in a 2:1 resonance while the outer planet is 
moved from the 2:1 to the 7:3 resonances with the third planet. 
Planet $e$ is started with
random orbital elements close to its expected orbit for a 2:1 and 5:2 
mean motion resonances
with planet $d$. 
The orbital elements are sampled randomly around this initial 
configuration looking for solutions which may be stable for long timespans. 

Assuming a mass for the star of 1.56 $M_{\odot}$ \citep{2011ApJ...741...55S} and 
for the planets the nominal masses 9,10,10,7 $M_{J}$
for $e,d,c,b$, respectively and by numerically integrating the 
orbits we find highly chaotic solutions that survive
on orbits similar to the observed ones 
for a few Myrs only. This is not compatible with the stellar age even assuming 
that the  primordial gas disk damped the eccentricities of the planets
granting stability. The disk lifetime can be 
no longer than $\sim 20 \times 10^6$ years. This is a conservative assumption as, 
according to observations, the presence of gas in the 
inner regions of the disk is observed only for timescales shorter than 
$10^7$ years \citep[see the review by][]{2011ARA&A..49...67W}. 
Of course, due to the chaotic nature of the orbits, it may be possible to 
find orbits surviving for longer timescales, as those found by \cite{2012arXiv1201.0561S},
but a different choice of the numerical integrator gives already different 
solutions undergoing close encounters after a few Myrs. 
As already noted by \citet{2010ApJ...710.1408F}, reducing
the planetary masses leads to longer timespans before the onset of 
a 'Jumping Jupiter' phase. 
By reducing all masses by 30\% and setting the outer planet in a 7:3 resonance
with the third planet and  planet $e$ in a 5:2 resonance with 
the second planet, we find the most stable solutions to the system,
according to the FMA analysis. They are still chaotic but they survive 
at least over the estimated stellar age. However, these solutions  
are not fully compatible with the observational data.
If the mass of the planets is reduced at 50\% of the original values, 
the phase space area where orbits survive for at least 30 Myr becomes significant
and even solutions with all four planets in mutual 2:1 resonance become 
stable over a longer timspan. 

While a 30\% reduction of the planetary masses is still compatible with the 
stellar age and theoretical models, a reduction of 50\%  
(about  3.5, 5, 5, and 5 $M_{J}$ for HR 8799 $b$,$c$,$d$,$e$ 
respectively) is below the current estimates.
Using \cite{2003A&A...402..701B} models, a stellar age of 15-20 Myr would 
be required to be compatible with such low masses for the planet, which is below
the youngest age derived by \citet{2010Natur.468.1080M} from membership to 
Columba association (nominal age 30 Myr)
and only marginally compatible with the youngest age estimate (20 Myr) by \cite{2006ApJ...644..525M}.
However, \citet{2008hsf2.book..757T} quoted a significant age uncertainty  for Columba association.
Furthermore, some age dispersion within the association or a small age difference between the
star and the planets might help to reconciliate evolutionary model predictions and 
constraints from dynamical stability of the system.
The $3 \sigma$ lower limits of the model atmospheres fit by \cite{2011ApJ...737...34M}
are 2, 6, and 3 $M_{J}$ and ages of 10, 20 and 10 Myr for planets $b$, $c$, and $d$, respectively.

Our explorative analysis can not be considered completed. For example, we did not explored the possible
impact of varying the stellar mass within the rather broad errorbars ($\pm0.3~M_{\odot}$) 
allowed by \cite{1999AJ....118.2993G} study. A full analysis is postponed to a forthcoming paper.

\section{Conclusion}
\label{s:conclusion}

We have performed H and Ks band observations using the new AO system at the Large Binocular Telescope
and the PISCES Camera. Analyses performed independently by different members of our team using two different 
pipelines yielded consistent results.
The excellent instrument performance (Strehl ratios up to 80\% in H band) enabled detection of the inner 
planet HR8799$e$ for the first time in the H band.
The H and Ks magnitudes of HR8799e are similar to those of planets $c$ and $d$, with planet $e$
slightly brighter. Therefore, the inner planet should have similar masses to $c$ and $d$, being likely slightly
more massive.
When considering only H/H$-$K and K/H$-$K color-magnitude diagrams, the positions of HR8799$c$,$d$, and $e$ 
is close to the field brown dwarf sequence, especially for the brightest planet $e$. 
Planet $b$ is instead significantly fainter than the other HR8799 planets, as already discussed in the literature.
The occurrence of some systematic differences at 0.2 mag level between photometric measurement of the HR8799 planets
from different sources have some impact on these conclusions and should be further investigated.

We also collected the available astrometric measurements of HR8799 planets, exploring
possible orbital configurations and their orbital stability.
We confirm that the orbits of planets $b$, $c$ and $e$ are consistent with being circular and coplanar;
planet $d$ should have either an orbital eccentricity of about 0.1 or be non-coplanar with respect to
$b$ and $c$. 
We found that the planet $e$ can not be in a circular and coplanar orbit with the other planets and in 4:2:1 mean motion 
resonances with planets $c$ and $d$. These resonances require significant deviations from coplanarity or
eccentricity. A coplanar and circular orbit with 5:2 resonance between $d$ and $e$ is instead 
compatible with the observational data.

We found the system to be highly unstable or chaotic when the nominal planetary masses are adopted. 
Significant regions of dynamical stability for timescales of tens of Myr are found only when adopting
planetary masses of about 3.5, 5, 5, and 5 $M_{J}$ for HR 8799 $b$, $c$, $d$, and $e$ respectively.
These masses are below the current estimates based on stellar age (30 Myr) and theoretical models
of substellar objects. A more complete exploration of the parameters space will be performed
in  a forthcoming study to identify possible system configurations that are compatible with the 
observations and dynamically stable. In any case it is more difficult to find dynamically stable
solutions for the 4-planet system than for the 3-planet case considered in most of the
literature studies. This implies smaller upper limits on planetary masses from 
dynamical stability constraints. These results will be relevant to place clues on the 
physical models of planet structure and atmospheres.

On the observational side, the continuation of the astrometric monitoring is mandatory for a better 
characterization of the system.
With an orbital period of about 50 yr for HR8799$e$, a few more years of observations will allow significant
refinement of the constraints we can put on the orbital parameters. New observations should pay specific
attention to achieve the best astrometric accuracy  (optimization of the observing procedure and instrument
set-up, dedicated astrometric calibrations, etc.).

\appendix
\section{Distortion correction coefficients}\label{app:distortion}

Drizzle coefficients for PISCES distortion and plate scale determination were obtained for drizzled and 
un-drizzled PISCES images from the sieve mask data.
For what concerns the fitted Drizzle coefficients, let us assume ${x', y'}$ are corrected centroid values in pixels, 
${x,y}$ are raw data centroid values, in 
pixels. Then ${x0, y0}$ translate the distortion equation to an appropriate centre for the distortion equation:

\begin{displaymath} x'=a_0+a_1(x-x0)+a_3(x-x0)^2+a_6(x-x0)^3+\end{displaymath}
\begin{displaymath} ~~~~~~   a_2(y-y0)+ a_5(x-x0)(y-y0)+a_7(x-x0)^2(y-y0)+ \end{displaymath}
\begin{displaymath} ~~~~~~   a_5(y-y0)^2+a_8(x-x0)(y-y0)^2+a_9(y-y0)^3  \end{displaymath}

\begin{displaymath} y'=b_0+b_1(x-x0)+b_3(x-x0)^2+b_6(x-x0)^3+\end{displaymath}
\begin{displaymath} ~~~~~~   b_2(y-y0)+b_5(x-x0)(y-y0)+b_7(x-x0)^2(y-y0)+  \end{displaymath}
\begin{displaymath} ~~~~~~   b_5(y-y0)^2+b_8(x-x0)(y-y0)^2+b_9(y-y0)^3  \end{displaymath}

The values of the a$_i$ and b$_i$  coefficients are listed in the Table~\ref{Tab:distortion}.

\begin{table*}
	\caption{Distortion coefficients for PISCES camera}
	\begin{center}
	\begin{tabular}{|l|c|c|c|c|c|c|c|c|c|c|}
	\hline
		a$_i$	& 0.313 & 0.999 & 6.84-4 &4.396e-6 & -4.266e-6 &  1.173e-6 & -5.777e-8 & -2.121e-9 & -6.117e-8 & 7.322e-10 \\ \hline
		b$_i$	& -0.010 & 6.116e-4 & 1.001 & -6.931 & 3.707e-6 & -6.027e-6 & -4.859e-10 & -6.436e-8 & -1.413e-9 & -5.796e-8 \\ \hline
	
	\end{tabular}
	\label{Tab:distortion}
	\end{center}
\end{table*}

\begin{acknowledgements}

This research has made use of the 
SIMBAD database, operated at CDS, Strasbourg, France. 
DM, RC, SD, FMar ackowledge support by INAF through 
PRIN-INAF 2010 ``Planetary systems at young ages''.
We thank Piero Salinari for his insight, leadership and persistence
which made the developement of the LBT adaptive secondaries possible.
We are grateful to Mr. Elliott Solheid, the lead mechanical engineer
on the adaptation of the PISCES camera to LBT AO system. 
Mr. Roland Sarlot and Mr. Andrew Rakich provided
support in optical design and engineering, respectively.
We warmly thank the anonymous referee for the helpful report.

\end{acknowledgements}

\bibliography{hr8799}
\bibliographystyle{aa}
           
\end{document}